# The impact of Facebook-Cambridge Analytica data scandal on the USA tech stock market: An event study based on clustering method

Vahidin Jeleskovic[*] and Yinan Wan[**]

**Abstract:**

This study delves into the intra-industry effects following a firm-specific scandal, with a particular focus on the Facebook data leakage scandal and its associated events within the U.S. tech industry and two additional relevant groups. We employ various metrics including daily spread, volatility, volume-weighted return, and CAPM-beta for the pre-analysis clustering, and subsequently utilize CAR (Cumulative Abnormal Return) to evaluate the impact on firms grouped within these clusters. From a broader industry viewpoint, significant positive CAARs are observed across U.S. sample firms over the three days post-scandal announcement, indicating no adverse impact on the tech sector overall. Conversely, after Facebook's initial quarterly earnings report, it showed a notable negative effect despite reported positive performance. The clustering principle should aid in identifying directly related companies and thus reducing the influence of randomness. This was indeed achieved for the effect of the key event, namely "The Effect of Congressional Hearing on Certain Clusters across U.S. Tech Stock Market," which was identified as delayed and significantly negative. Therefore, we recommend applying the clustering method when conducting such or similar event studies.

## 1. Introduction

With Internet technology achieving brilliant achievements nowadays, we are also using social media more widely in our daily life. However, though the high-speed development of technology has made our lives more convenient, it also poses a threat to the online privacy of Internet users. The most dramatic event at the beginning of 2018, would be the unprecedented disaster encountered by Facebook. A political consulting and strategic communication firm, named Cambridge Analytica, developed a personality test application that obtained personal information not only from participants, but also from their friends in their social accounts, illicitly using the personal data of up to 87 million Facebook users. More importantly, all the data and information were obtained through a partnership with Facebook. As a result, Facebook was accused of mismanaging user information. On March 22, 2018, Facebook chairman and chief executive officer Mark Zuckerberg issued an apology and admitted that the company did not protect the users' data. He said a series of strict measures would be taken to ensure that similar incidents do not happen again (MEIXLER, 2018). According to Zuckerberg's declaration at a Congressional hearing (Deutsche Welle, 2018), we can find that the core of data leakage focuses mainly on third-party application platforms. Although Facebook has strengthened the management of its user data and drawn up the regulations with third-party platforms, Aleksandr Kogan still shared the private data with Cambridge Analytica, which was said to push targeted ads, particularly political advertising. These influenced Donald Trump's 2016 presidential campaign and other major political events. It is obvious that the ads on Facebook will directly influence users' expression, information seeking and real-world voting behavior (Bond et al., 2012). As a result, Facebook was exposed to the most serious crisis caused by these data breaches and suffered several serious consequences.

---

[*] Humboldt-Universität zu Berlin, Spandauer Str. 1, D-10178 Berlin. Email: vahidin.jeleskovic@hu-berlin.de
[**] Independent Researcher. Email: yinyang.yinanwan@gmail.com

A common piece of advice from a variety of researchers on online safety is that users should enhance their sense of personal responsibility and take steps to protect themselves from the online attacks (Shillair et al., 2015). However, there has been little previous evidence of the economic impact of such data leakage in the field of information security. Therefore, in this paper, we will mainly focus on the impact on the stock market, particularly on the U.S. stock market caused by a scandal related to the misuse of users' personal data. It will be interesting to know whether this kind of corporate scandal is still correlated with stock market reactions. Broadly speaking, numerous studies have investigated the effect of a firm-specific corporate event on the affected firm's securities by using a method called event studies (Carlisle & Patton, 2013). The aim of this paper is to provide an empirical evaluation of the recent data leakage scandal concerning Facebook, to confirm whether there is any overall impact on other industry-related companies. It has significant benefits in terms of data monitoring on the third-party developer. Furthermore, Facebook is the world's largest social platform, with its core assets of billions of units of data. It is expected to take the responsibility to protect the private data of users when it is also a corporate social responsibility (CSR) (Janney & Gove, 2011). It was also a cause for alarm at the General Data Protection Regulation (GDPR) in the European Union.

In order to concretize this approach, this paper will proceed as follows. Firstly, we will undertake a brief description of the methods we will use in section 3. Also, we will put forward the relevant hypothesis. Secondly, in section 4, we will make sure of the date on which the scandal first exposed. Moreover, on account of this scandal, we will analyze the overall impact of further events across the sample firms, while all the further events are Facebook-specific and might be concerned with the data leakage event. At the same time, it is also necessary to determine the day of announcement for these further events. In addition, previous work has been unable to follow the changing samples after each event. Therefore, our research will also examine how the samples change after the event by using some kinds of variables based on cluster analysis to achieve further insights into the reaction for such samples. The portfolios formed by clustering stocks, which contain Facebook Inc. can be seen as our subsamples. At the same time, we will measure the daily expected return by using the Market Model (MacKinlay, 1997) in this section. The abnormal return for each stock can be calculated by taking the difference between the return of a single stock and the return of a reference portfolio (Warner & Brown, 1980), where the NASDAQ 100 index is the reference portfolio. Following the methodology developed by Black, Jensen, and Scholes (1972), we will then test our hypotheses by examining the cumulative average abnormal return (CAAR), which will be further explained in the next sections, in cross-sectional variance across all the sample firms, and to make sure whether the cumulative average abnormal return is significantly different from zero. While many researchers have investigate the impact on an industry, there is much less empirical research discussing how a firm-specific event has influenced the effect on a firm's own stock. As such, after examining the overall impact of the earnings report for the second quarter across the sample firms, we wonder what effect this report has had on Facebook itself, which is obvious since the equity would probably react immediately. All the results will be described in section 5. Finally, we will provide a conclusion, along with a description of limitation at the end of this paper.

## 2. Literature Review

Contrary to the literature on the analysis of the economics of personal crime (Strachan & Beedles, 1983), research on corporate wrongdoing has recently become a major issue. As Utz (2019) defined, a scandal can be a publicly unknown weakness in the company, when the information about it is made public, it will cause a widespread controversy. These kind of scandals are mainly regarded as unethical or even illegal activity (Long & Rao, 1995). An unethical activity, or even the illegal acts are defined as illegal payments, employee discrimination, white-collar crime (Tay, Puah, Brahmana, & Abdul Malek, 2016), restatement of a financial earning (Robbani & Bhuyan, 2010), product recalls (Davidson & Worrell 1992; Jarrel & Peltzman 1985), or insider trading (Long and Rao 1995). Besides these kind of unanticipated unethical performance, many researchers consider the anticipated or partially anticipated event like central bank intervention (Trivedi & Srinivasan, 2016), stock dividend announcement (Anderson, 2009), corporate acquisitions (Malatesta & Thompson, 1985), public policy (A. J. Thompson, 1993), bankruptcy announcement (Lang & Stulz, 1992; Aharony & Swary, 1983). Furthermore, ecological and humanitarian disasters such as nuclear accident (Hill & Schneeweis, 1983; Ferstl, Utz, & Wimmer, 2011) are also take into consideration.

Prior studies on the impact of corporate scandal on stock price were mainly focus on the release of various information by listed firms. The most attractive one is for the restatement of a financial earning (Robbani & Bhuyan, 2010). DeFond and Jiambalvo (1991) study the incidences and circumstances of accounting errors made by 41 firms. Their analysis indicates that smaller and less profitable firms are more likely to restate the financial result than larger and stable companies, which is consistent with the result found by Kinney and McDaniel (1989), who analyze 178 quarterly earnings errors from 73 firms. The negative CARs of the entire industry portfolio caused by firms restating its quarterly earnings indicate that either overstatement or understatement will impact the stock market.

Many researchers believe that there is a certain relationship between the announcement of bad new and company performance. In fact, if directors and other insiders are likely to obtain abnormal profits if they already receive the information before its published (Park, Jang, & Loeb, 1995). As a consequence, insider trading refers to legal as well as illegal transaction under U.S. law, whereas it is generally exclusively refers to illegal trading behavior. Allen and Ramanan (1995) measure the cumulative abnormal return over a 15-month period from the beginning of the fiscal year because the insider trading information will be reflected on the price more than six months. From the five categories of the insider trading they classified, it is obvious that insider trading can convey important information regarding the stock price of a firm's equity in the future, which is widely known as well (Ball & Brown, 1968). What's more, there is a positive unexpected earnings in Max buy trading. On the contrast, the negative unexpected returns are resulted in unfavorable information. This result is also confirmed by Seyhun and Bradley (1997), which conduct 525 bankrupt firms between 1963 and 1992.

Empirical studies which have a viewpoint that illegal backdating behavior will have negative cumulative returns is outlined by Long and Rao (1995), who have found that there is a negative CAR statistic at 1% level. Tay et al. (2016) report in their findings a negative reaction from Malaysian-listed companies regarding to the announcement of white-collar crime. In support of the white-collar crime, Hilb (2008, pp.29) described a corporate accounting scandal from ENRON, which indicated that no matter how successful you were, it would be experienced serious stock market collapses without integrity. When Andersen's reputation was negatively affected by this audit failure event, Chaney and Philipich (2002) employ event study with 3 dividend categories and suggest that all Andersen clients, especially the Houston clients, make a negative reaction in the following two days around the event. Linthicum, Reitenga, & Sanchez (2010) argue how a firm value affected by corporate social responsibility during the Enron crisis. In contrast to most research, which report that the high level of social responsibility in the enterprise will be beneficial in the event crisis (Grow et al., 2005), they claim that even the great social responsibility can not mitigated the negative returns to Anderson's client.

Another same conclusion is supported by the study of a product-recall scandal, which is correlated with corporate social responsibility and reputation theory (Janney & Gove, 2011). Davidson III and Worrell (1992) examine the reaction of the stock market to non-automobile recalls in industry, rather than in automobile industry. As they mentioned, there are diversity of recall types. For example, a product can either be replaced or be repaired or even refunded. Besides, manufacturers can volunteer to recall a product or be asked by the government. Due to the different types of recalls contain different information so that the stock market is more likely to response differently. According to their research, it shows that it is more negative when a product is replaced or return the purchase price compared to repair the product. Second, the stock market reacts more negatively to those government-ordered recalls than those voluntary recalls. Meanwhile, when a product taken off the market, it is also a worse signal to the stock market than a normal recall. Since the greater complexity of products as well as the higher requirements for product quality and safety, product recalls are probably to occur more frequently in the future (Berman, 1999). One of the most famous recall in recent year was the Volkswagen emission scandal in 2014 (Bachmann ey al., 2017). The financial valuations in both BMW and Mercedes-Benz are suffered by scandal because of the collective reputation as German manufacturers. That is, the collective reputation (Tirole, 1996) definitely suffered great loss due to the scandal because corporate reputation is closely related to product quality (Waddock & Graves, 1997). Moreover, the scandal was widely covered in the media, which lead to a substantial reduction not only in Volkswagen's stock price, but also in the non-VW German auto manufacturers. An event study is used to detect that both BMW and Daimler suffered a negative cumulative abnormal return. Nevertheless, Mitchell and Stafford (2000) argue that event studies may have some limitations in examining long-term effects.

As already been shown, many of the researches are focusing on the overall industry-widely impact from the scandal. King (1966) hypothesizes that prices tend to move together for companies in the same industry. Even the effect of the ecological disaster like Three Mile Inland accident on bond risk premia to electric utility firms as a whole (Barrett, Heuson, & Kolb, 1986). Thus, when a scandal exposure in a firm, the other firms in a same industry have negative reaction caused by such problem, or even be re-evaluated, can be regarded as a contagion effect (Aharony & Swary, 1983). In addition to considering the impacts of companies with similar characteristic, they also measure the impact on competitors who attract more investors during rival's scandal period (Jarrell & Peltzman, 1985). As they mentioned, as one of the Detroit "Big Three" automakers, a recall from Chrysler will not have a great impact on his competitor, Ford and GM. Lang and Stulz (1992) call it competitive effect of an announcement.

On the other hand, it is hard to say how exactly an economic behavior influence a industry. In most cases, there would be both contagion and competitive effect for their competitors in the same industry. The impact also depends on the type of industry due to the different characteristics, such as the degree of concentration (Lang & Stulz, 1992) for the firm and industry. Moreover, the extent of an impact is related to such indicators like stock yield, financial leverage ratio, size, book-to-market ratio, cash flow

(Ikenberry, Lakonishok, & Vermaelen, 1995; Tawatnuntachai & D'Mello, 2002). Based on these findings, it is possible to formulate the hypothesis:

H0-1: the data leakage scandal from Facebook had no effect on stock price across the US tech-stocks market.

When Brooks, Patel, and Su (2003) argue that after the announcement, selling pressure, trading volume, and quoted dollar spreads will increase, they choose five variables, which are stock price, spread, volume, volatility and trading location into the regression to examine if the event has a contagion effect on companies in the same industry. In addition, Amihud and Mendelson (1986) group stocks based on their spread and relative risks to form a portfolio. This is consistent with the arguments that the firm with similar characteristic will have more contagion effect (Lang & Stulz, 1992). To determine, if there is a difference between the portfolios with some particular variables, we further expect to hypothesize that:

H0-2: the data leakage scandal from Facebook had no effect on all clusters based on particular risk measures in the US tech-stocks market.

## 3. Methedology

Primarily, we focus on an available measurement to support the financial effect of a corporate event on our sample securities. Following various of literature, we employ an event study method to investigate the performance of stock price. Generally speaking, a market model will be concerned in a semi strong form test (Fama, 1970). In order to create a more precise sample groups, which are similar to each other, we can use cluster analysis with some chosen variables.

### 3.1. Event Study

Nowadays, event studies have become the predominant methodology to examine the impacts of an unanticipated event on the capital market. The theoretical basis of the event study is the semi-strong form of the market efficient hypothesis by Fama (1970). Generally speaking, the capital market is always assumed to be efficient and rational (Fama, 1965). The impact of the event can be immediately reflected in the price of the security. That is, the degree of deviation of the average excess return or, cumulative excess return from zero in a certain event window can be determined by using the event study method. In short, companies' announcements provide market participants with information that can be included in the market price (Konchitchki & O'Leary, 2011). We therefore measure the difference of stock return called abnormal return, in a specific time period before and after the announcement so as to find out how it affects the abnormal return.

The expected return on the event day can be calculated by for example, statistical models, which include constant mean return model, market-adjusted return model, market model (MacKinlay, 1997), multi-factor models (Fama & French, 1992, 1993) and so on. By simulating the daily return by U.S. NYSE/ASE, Brown and Warner (1980) found that when we use simple market-adjust return model and market model, the result are always satisfied in some circumstances. They also pointed out that though the different characteristics of different models can be compared by mathematic deduction, the discrepancy between them cannot be reflected in an empirical test. Hence, so as to ensure the precision of a study and to improve the choice of the models, it is necessary to undertake a comparison between all these models at different times and in different markets. While we do not discuss all the models in this paper, we estimate the abnormal return as regards the estimation procedure, adopting a standard market model, which will be explained in the following part. Apart from statistical models, we will also take economic models into consideration. However, Economic models restrict the parameters of statistical models to provide more constrained normal return models. Most important of these are capital asset pricing model (Black et al., 1972), and arbitrage pricing model (Solnik, 1982).

In fact, Dolley (1933) was the first to use event study to investigate the returns effect of stock split by using 95 stock split-up events from 1921 to 1931 in the U.S. stock market. He found that 57 of them induced rising prices, 26 of them decreasing prices, and the remaining 12 had almost no reaction to this event (Dolley, 1933). Since Fama, Fisher, Jensen, and Roll (1969) further improve this method, they attempt to determine the effect of split when the dividends increase is excluded. Meanwhile, the modern methodology of event studies was confirmed. That is, we can conclude that the work of FFJR marked the beginning of a new area in the capital markets. By the 1980s, a large number of papers use this method to study wealth acquisition appear. The approach has been modified to resolve the statistical issues so as to make event study statistically valid at the same time. Brown and Warner (1980) use monthly return in event study, while it was well-specified with market model under various conditions. Daily data was described in their further research in 1985 (Brown & Warner, 1985). McWilliams and Siegel (1997) re-examine the three recent studies with cross-sectional independence test in the light of Patell's (1976) test procedure, which widely popularized the utilization of standard abnormal returns. Subsequently, in an effort to resolve some econometric issues in event study, Kolari and Pynnönen (2010) provide a more powerful test method. It is now widely used in various disciplines, such as business, accounting, finance, organizational behavior and the various subfields of economics as well, especially regulatory economics (Binder, 1998). The most successful area of application of the event study is in corporate finance since the research objects in this field focus on the changes of abnormal earnings before and after the first announcement of listed companies. For example, announcements of acquisitions, stock splits, mergers, annual reports, and new security issues. More recently, according to Agrawal and Kamakura (1995),the methodology can also be used to evaluate how marketing-related events affect a firm's profitability, such as changing a company's name (Horsky & Swyngedouw, 1987), recalling product (Davidson & Worrell, 1992), brand leveraging (Lane & Jacobson 1995) as well. In addition, Filbeck, Gorman, and Greenlee (2005) investigate on supply chain technology, software, and systems, so as to reach a desired goal which can be higher profits and maximize shareholders value, that is, whether they can have affect the stock market by a given company.

Frequently, the process of the event study, as MacKinlay (1997) outlined, involves the following steps: 1). Identifying an interested event; 2). Defining the event window; 3). Determining the selection criteria to set up firms as samples in analysis; 4). Predicting normal returns throughout the event window; 5). Finding out the statistical properties of the abnormal return; 6). Testing whether the abnormal return has a statistical difference from zero.

**Market Model**

In an endeavor to assess the relationship between daily return from Facebook and the market portfolio, we can make a regression analysis, in which the independent variable is the daily return from NASDAQ100 and the dependent variable is the daily return from Facebook. In general, ordinary least squares can solve the market model parameters in a good manner (Brown & Warner, 1985). It is the most popularly employed model in event studies, representing a potential improvement relative to the constant mean return model (Strong, 1992).

Let $R_{i,t}$ be designated as the expected return in the period t, and the most important OLS to calculate the α and β. These two parameters are estimated over a clean estimation period. For any security i, the market model is:

$$R_{it} = \alpha_i + \beta_i R_{mt} + \varepsilon_i \tag{3.1.1}$$

where:

$$\beta_i = \frac{\text{cov}(R_i, R_M)}{\sigma^2(R_M)} \qquad \alpha_i = R_{it} - \beta_i R_m$$

$R_{it}$ is expected return of stock i on day t,

$R_{mt}$ is the market return on day t,

$\alpha_i$ is intercept term,

$\beta_i$ is security beta, the sensitivity of a stock to market return,

$\varepsilon_i$ is the prediction error, with $E(\varepsilon_i)=0$.

## 3.2. Cluster Analysis

Since data mining is becoming an emerging technology, it is related to the database, and is typically connected to the logic, statistics, machine learning, fuzziology, or use of visual calculating. A vast quantity of taxonomic techniques has long existed in life science whereas a rather important class of clustering techniques has been developed by researchers in behavioral disciplines in business firms as well (MacKinlay, 1997). Cluster analysis plays an important role in data mining. It is a family of techniques that allows the researchers to systematically identify similar entities across a number of different characteristics (Hambrick, 1983), in which refers to a group of multivariate mathematical algorithms. Accordingly, observations that scored in a similar manner by multiple specified variables can be found in the same group. However, the objects in different groups are perceived by the analyst with more differences. This method helps aggregate the categories when we do not have any prior knowledge for the belonging of each entity (Green, Frank, & Robinson, 1967). Because of the significance and particularity of this method, it has seen great development in recent years, with several cluster methods springing up, such as the partitioning method, hierarchical method, density-based, grid-based, model-based cluster analysis and so on (Pilevar & Sukumar, 2005).

Essentially, we consider the cluster would be changed in different time periods during each event. To illustrate this— the impact of the scandal on different Facebook-based clusters in different time periods, we use the hierarchical method with five chosen variables in the following section. As we use a hierarchical tree to join our samples together, we have to use distances between our samples when forming the clusters (Green et al., 1967). That is, the distance of each firm can be changed with different variables. According to Cavusgil, Chan, and Zhang (2003), our paper corresponds to Squared Euclidean Distance (dist = $\sum_{i=1}^{n}(x_i - y_i)^2$) ), which is a most commonly choice type of distance. Because of the square distance, a progressive larger differences in distance calculations are taken into account, which leads to the objects further apart (Gauch & Whittakert, 1981). Apart from these methods, a normal Euclidean Distance (dist= $\sqrt{\sum_{i=1}^{n}(x_i - y_i)^2}$), a sum of absolute differences between value pairs called Block/Manhattan Distance (dist = $\sum_{i=1}^{n}|x_i - y_i|$)) and Minkowski Distance (dist = $\sum_{i=1}^{n}|x_i - y_i|^r)^{1/r}$) ) can be used as well.

## 4. Experimental findings

### 4.1. Choice of the Window

For purpose of estimating the normal and abnormal returns, 'windows' are necessary in this event, for example, event window and estimation window. As shown before, four major events will be mentioned in this paper. That means, we will have four major event days (see Table 1). An event day should, therefore, be the first day when the news announced through social media.

| Facebook data privacy scandal | Congress grills Facebook's CEO over data misuse | Facebook reports first quarter earnings; Graph API and other platform product changes | Facebook reports second quarter results |
|---|---|---|---|
| 19.03.2018 | 10.04.2018 | 26.04.2018 | 26.07.2018 |

Table 1: Major Event Days

In the next step, we examine a pre- and post-announcement window, which comprises the days around an event date, to check out whether the event influence the performance of the firm's stock. The time interval can be one day, one week, or even a few months. But in essence, the use of a long window stacked the deck against the result which had been hypothesized (Meznar, Nigh, & Kwok, 1998). They also found that there would be greater possibility to meet other events during the window and confound or obscure the impact of the announcement on the stock returns when we examine a longer window. Various lengths of such kind of windows have been chosen in the past researches. For example, Barber and Lyon (1997) test the abnormal stock returns in a long-run period with annual, three-year, and five-year. This is in contrast to WRIGHT and FERRIS (1997), in which a window is reported with 21 days, consisting with day -10 to -1, the event day 0, and day +1 to +10. Apart from studying several periods between –20 to 0, -30 to 10, -20 to 10, -10 to 10 and -2 to 10 periods, Meznar et al. (1998) show a closer window over -2 to 0, -1, to 0, -5 to 5 windows, in order to find a more moderate association between announcement and share price based on their previous study. Our study also focuses on the period around the announcement day, which will be investigated with the 5 days before and 20 days, if possible, after the announcement day. The reason for this short prior window is because that we have to reduce the impact from additional noisy news, or even by Facebook itself, which interfere with the market's reaction (Brown & Warner, 1985). We also focus on a differing window lengths ranging from six days, from day 0 to day +5 (Meznar et al., 1998), and aim to make a short observation of a week-impact on the industry. Our detailed windows are denoted as follows:

Event 1: On March 17, 2018, it was reported by *The New York Times* (Isaac & Frenkel, 2018), originally reported that Cambridge Analytica collected 50 million Facebook profiles, this number was later revised to "up to 87 million" profiles. Two days later, on March. 19, when the stock market reopened on Monday, the stock price fell sharply by almost 6.77%. When such an event occurs overnight, investors have no chance to trade on the information. After a period of no trading activities, the price will respond immediately to the market (Brooks et al., 2003). Therefore, we define March 19, 2018 as our first event day, denoted as t=0. Specifically, when the event was suddenly disclosed by *The New York Times*, we consider that an emergency. What is more, we have not found any other prior news about the upcoming contract for the event. Although we assume that there was no information leakage before the event, we still make a simple test about the pre-announcement period. As such, the corresponding pre-announcement window is (-5, -1) and the post-announcement window is (0, +14) in this event, in which the day +14 is the day before the second event day.

Event 2: When Zuckerberg decided to testify before Congress, Facebook suffered a great pressure in the wake of Cambridge Analytica scandal (Deutsche Welle). About two weeks later, after Facebook's CEO already took out a full page to apologize to consumers for not properly securing their personal data, he testified in front of the U.S. Congress for two days on this issue. Define April 10, 2018 as day 0, and the surrounding 17 days (-5, -1) and (0, +11) are designated the pre-announcement window and post-announcement window.

Event 3: On late April 26, 2018, Facebook reported its first quarter earnings. The better-than-expected results led to a surge in the stock market on the following day—April 26, 2018, which put Facebook on pace for its best day since January 2016. We, therefore, define April 26, 2018 as the event day, while the surrounding 26 days (-5, -1) and (0, +20) are the pre- and post-announcement windows.

Event 4: On July 26, 2018, the trading day together with the announcement of the second quarter report. The Facebook stock plummeted after the earnings report. This disappointing report was regarded the largest one-day destruction since it went public, wiping out roughly $120 billion of shareholder wealth. It warned that revenue would grow slowly in coming quarters, and the company would invest heavily in privacy and security. Furthermore, it is quite possible that the investors showed a negative emotion since Facebook suffered the data scandal, along with online abuse, government probes, and predicted a worse expectation on this scandal-related quarterly earnings. Therefore, we would like to define the day as our event day 0. As a result, we examine the 26-day period from -5 to -1 and 0 to +20 as our pre- and post-announcement windows.

Consistent with most of the existing research, we will concentrate on establishing an estimation window. The most commonly technique is to take the period before the event window as the estimation window (MacKinlay, 1997). There is no coercive provision about the length to the 'windows'. Following the previous studies, Park (2004) conduct an analysis with several estimation windows, which are 100, 150, 200 and 250 days, and find out that no matter how long the window takes, the abnormal returns are almost the same on the event day with the market model. Lee and Connolly (2010) regress the return in accordance with the market portfolio for 200 business days, from day -210 to day -10, in order to investigate how different kinds of technology news affect firm value. The majority of authors recommended that we should choose at least 100 days as our estimation window in order to make sure that the chosen windows are clean, with no event-affected factors. On the other hand, Rozeff and Kinney (1976) suggest that January is always the beginning of the tax year for investors and of tax and fiscal years for most of the firms. Keim (1983) shows evidence that the January mean daily excess returns are greater relative to other months, especially small firms suffer much of the excess return in January. More than 50 percent of huge abnormal returns are demonstrated in the first trading week of January, especially on the first trading day. He explained the phenomenon as the 'January effect'. In this paper, we take this anticipated fiscal year-end into consideration and set up the parameter estimation period starting at day -52 and ending at day -1 relative to the first event (Fama & French, 2015; Warner & Brown, 1980), which is on March 19, 2018, when Facebook first announced that it had failed to safeguard user data. The prior 52 days in this period ($T_0$) are designated as the estimation window because these 52 days are just the one and the same state of the market. After the day 0, we have another state of the market. Regarding the following three events, which are all the follow-on events, we consider the same estimation window as the first event.

**4.2. Sample Construction**

Our goal is to develop analytical classification techniques that differ from most existing researches, and are classified systematically, directly by an industry. We will describe the data collection procedure and how the samples are classified in different clusters. In this paper, firms were selected in accordance with the following procedure. First, the appropriate index for event studies is an issue. As we all know, NASDAQ100 (NDX) is a modified capitalization-weighted index and is always an efficient market as well. These firms are made up of the 100 greatest non-financial local and international listed companies' stock indexes as exchange traded funds, especially larger market-valued growth stocks. As one of the most important indexes in NASDAQ, NDX has the characteristics of being high-tech, high-growth, and non-financial. It can arguably can be seen as the representative of the U.S. tech stocks. The majority among the indexes belongs to the technology industry, in which Apple accounts for a large weight, while Microsoft, Google, Intel, and our research object Facebook are also included. The applied index construction can be obtained in different ways. Warner and Brown (1980) use Center for Research in Security Prices at University of Chicago (CRSP) to get the available daily return data, when Duso, Gugler, and Yurtoglu (2010) collect stock data by using DataStream. In our paper, we choose value-weighted portfolio of NDX as the market index in Thomson Reuters DataStream to estimate the daily stock returns, which is our primary data source. The $R^2$ thus regarded is adjusted for degrees of freedom. In the words of MacKinlay (1997), "the higher the $R^2$, the greater is the variance reduction of the abnormal return, and the larger is the gain". This is the reason why we choose NDX as our market index. When the index has been confirmed, it comes to focus on the daily stock price over the period from 01.01.2018 to 31.08.2018. Only firms in United

States were chosen, which contains almost all the securities in this equity index. Apart from 'Software and Computer Services', the industry called 'Technology Hardware and Equipment' is also considered a valid observations, while 16 are in 'Software and Computer Service' and 21 in 'Technology Hardware and Equipment' respectively. We, therefore, consider our initial samples as being representative of the US tech stock market. We then obtain corresponding stock data and daily prices on the NDX index and on 37 other securities as well.

Previous research measured the trading process after announcements of an unanticipated event by several variables. Given that the stock markets is very different in terms of trading volume, or other risk-related factors, suppose the magnitude of response is therefore depends on firm size and differences in risks (De Bondt & Thaler, 1987). King (1966) relates factor analysis to covariance matrix of monthly changes in closing prices. Another particular multivariate analysis is industrial clustering in subgroups that provides a rough indication of the relevant within the cluster. Fleming and Remolona (1999) illuminate that trading volume and spreads react immediately when macroeconomic news is released. Foster and Viswanathan (1993), for example, rely on the stock market trading volume, trading cost, and return volatility in view of the microstructure of trading between different traders, and advice that the intra-day trading volume will be very high when the returns volatilize at a high level, both before and following the announcement. Brooks et al. (2003) look at a series of unexpected events and verify the market response around the time period to those unanticipated negative events by examining price, trading volume, spreads, and trading location. Since the last two events for earnings reports are probably anticipated, spreads, volume, and volatility are likely to change during the pre-announcement period.

Given that the variables react immediately after the event, we would like to delineate our samples, as we are more convinced with the factors prior to each event. In a similar manner, the previously selected 37 entities are to be categorized on the basis of 5 chosen variables, which are identified as the risks for trading activities. But unlike most researchers, we examine the firm-specific characteristics among different entities in order to seek hidden similarities, which are not immediately obvious, rather than clarify how the variables affect the trading activity (Greene & Watts, 1996). Cluster analysis will be used to categorize the similar entities into a strategic group based upon the variables measured. In particular, we should determine the difference between each event. That means each cluster before and after the event will be taken into consideration.

Many researchers investigate trading-period return into the minute (Harris, 1986; Brooks et al., 2003) or across an trading hour (Foster & Viswanathan, 1993). Hillier and Marshall (2002) proposed a view that there would be more trades from the insiders during 20 days before and after the announcement than during any other period. We, therefore, use 20 days prior to the announcement as the time period to do the clustering. If, for example in our case, a following announcement of the same stock released immediately, the trading days before event 2 and event 3 is not that long, we conclude all days before these two events as our observed time period, which are 14 and 11 days respectively. Hence, the post-announcement period of the previous event disclosure also constitute the pre-announcement period of the subsequent event. Similarly, in order to find out the difference after the last event, we should also do the clustering with 20 days after the final chosen event.

Ding (1999) analyzes time-series variations of the spread in the foreign exchange futures market. As Chordia, Sarkar, and Subrahmanyam (2005) focus on the market spread depths, and trading activity over an extended time sample, both quote the effective spread, which they investigate, and which increased dramatically in down markets. Krinsky and Lee (1996) delineate an expansion of spread before earnings announcement. They show both effective spread and quoted spread. They also show that the average absolute daily variance in the spread is smaller than the average daily change in volume and other variables.

When the most researchers concentrate on the bid-ask spread (Härdle, Hautsch, & Overbeck, 2009, P379), here, we quote the daily spread associated with the both highest and lowest price for each trading day. In order to examine the price variation over a defined period of time, we calculate the difference for each day and get the average value for the selected 37 entities (see Equation 4.2.1).

$$daily\ spread = \frac{1}{N}\sum_{n=1}^{N}(P_h - P_l) \qquad (4.2.1)$$

where: $P_h$ is the high price for a trading day;

$p_l$ is the low price for a trading day;

N is the number of observed days.

Stock price and volatility are the direct response to the stock market. These are closely related to financial risks because the trend of fluctuation reflects the trend of financial risk. The variance and covariance between portfolio risk factors determine the volatility of portfolio returns (Alexander, 1998). Many researchers study changes in aggregate volatility in stock market. Officer (1973) relates the business fluctuations to market-factor variability over the period of 1897 to 1969. In fact, many researchers estimate the price variation over a defined period of time by using the high, low, opening, and closing prices and the transaction volume (Garman & Klass, 1980; Rogers & Satchell, 1991), because these prices comprise information that is readily available. Further, this variance of the logarithm of original price is called volatility (Garman & Klass, 1980). Generally speaking, the return in stock market always shows like that: the huge fluctuation follows a larger volatility and a small volatility follows on a weak fluctuation. The trend of fluctuation reflects the uncertainty and risk trend of finance. Furthermore, it has also the leverage effect (Schwert, 1989), which is a relatively small part of stock volatility. One explanation for the volatility of time-varying stocks is that leverage ratios change with relative stock and bond price. To model conditional volatility, a parametric econometric models like autoregressive conditional heteroscedasticity, which was first promoted by Engle (1982), can be used. An extension of this approach named generalized autoregressive conditional heteroscedasticity (GARCH) allows the method to support changed over the time dependent volatility. Nikkinen, Omran, Sahlström, and Äijö (2006) examine the behavior of GARCH volatilities in the response of global stock markets to the U.S. macroeconomic news, about 10 important news items in six regions, with a time-varying parameter. As Glosten, Jagannathan, and Runkle (1993) suggested, both unexpected positive excess returns and negative excess returns of stocks will change the conditional volatility of stock excess returns in the next period. As we mentioned earlier, the period after the first event is the same period before the next event. Usually volatility and correlation valuations are on account of daily or intra-day returns, as even weekly data may miss some of the turbulence in financial markets. We, therefore, estimate the volatility here as the sum of the squared daily returns (Schwert, 1989).

$$V = \sum_{i=1}^{N_t} R_{i,t}^2 \qquad (4.2.2)$$

According to Hiemstra and Jones (1994), daily return on stock is a continuous rate of of return in time-series analysis. Therefore, we define the daily stock return between times t and t-1 as:

$$R_{i,t} = \ln\left(\frac{P_{i,t}}{P_{i,t-1}}\right) \qquad (4.2.3)$$

while: $P_t$ means price on asset i on day t

$P_{t-1}$ means price on asset i on day t-1

$R_i$ means log-return on day t

In the third alternative version we check for the influence on volumes and prices to assess market evolution. Many researchers found that the trading volume is always important for observed stock returns patterns. They will especially increase or decline especially around an announcement (Keim, 1983). In addition, it will be different in hours of the day and a few days of the week. What is more, they conclude that the positive returns suffer steeper relationship between trading volume and returns than for non-positive returns (Jain & Joh, 1988). Others also find evidence that there is a nonlinear causality relationship between both volume and stock returns over several years (Hiemstra & Jones, 1994). However, while the firm size of the technology industry sample companies is totally different, the availability of information may increase the familiarity of viewers with their activities, thereby benefiting large companies (Tversky & Kahneman, 1974). Yet, it seems the trading volume is much larger in larger firms. That is, by using volume direct into clustering, it seems that the result will completely clustered in accordance with this variable when others are make no sense. Considering that there is a large relationship between volume and total trading volume, the volume weighted return as a daily representative return can better estimate of the market evolution (Lucia & Manuel, 2007). To this end, we used a market strength-weighted return, which is one of the specifications for volume-weighted return given by Lucia and Manuel (2007):

$$RVW = \sum_{t=1}^{T} \frac{v_t}{V} \ln(P_{i,t}/P_{i,t-1}) 100 \qquad (4.2.4)$$

where: $v_t$ is the trading volume on the day t

$V$ is the total trading volume during the particular time period

$P_t$ means price on asset i on day t

$P_{t-1}$ means price on asset i on day t-1

Apart from above mentioned variables, another risk remains, known as systematic risk, which is called beta, and which is used to compensate investors for the risk they take. Generally speaking, there are two ways to measure the stock risk, one is volatility and the other is relative to market. It represents the sensitivity of financial assets to overall market movements (Härdle et al., 2009, P285), which is a central parameter in modern finance of security returns and is now widely used to measure the asset's systematic risk (Black et al., 1972). The parameter beta will continuously increase or decrease after an event. A high beta coefficient means that macroeconomic changes are able to greatly affect assets, so that the variance is large. A low beta coefficient means that market changes are not able to the asset is not seriously affected by, hence, the expected return may also be lower (Kampman, 2011). Size-related differences in average return are explained by beta as well (Chan & Chen, 1988). Empirical results suggest that the beta is time-varying rather than constant (Bar-Yosef & Brown, 1977). Furthermore, a study by Brennan and Copeland (1988) found that the beta increased by nearly 20%, which is from 1.04 to 1.30, after the split-announcement day. That means, stock split and stock dividend may be accompanied by high uncertainty of future dividend, which result in the increase in systematic risk.

Large amount of researches have, therefore, been devoted to CAPM betas (De Bondt & Thaler, 1987). Investors make a decision to rely on the expected return and risk and, when evaluating the portfolio, they also rely on the expected return, variance and standard variance. According to the general form of CAPM model (see Equation 4.2.5), we use Mkt-RF for daily returns in Fama-French North American 3 Factors from the Data Library during each time period, and the other 37 securities are listed as the stock return. Hence, we can use Ordinary Least Square (OLS) to estimate the regression equation so that the alpha and beta can be measured as well.

$$R_{it} - R_{ft} = \alpha_i + \beta_i(R_{mt} - R_{ft}) + \varepsilon_{it} \qquad (4.2.5)$$

where: $R_{it}$ is the return on the investment portfolio i on day t,

$R_{ft}$ is the risk-free rate on day t,

$R_{mt}$ is the return on market portfolio on day t,

$\alpha_i$ is the intercept,

$\varepsilon i$ is the prediction error, and,

$R_{it}$-$R_{ft}$ is the market risk premium

$\beta_i$ is the regression coefficient on risk market portfolio of asset i, and is the covariance of the security return with the market return divided by the variance of market return, which is defined as:

$$\beta_i = \frac{cov\ (R_i, R_M)}{\sigma^2(R_M)} \qquad (4.2.6)$$

Aside from the determination of these factors, we finally pay attention to our final variable, CARs in our observed period. As Brown and Warner (1985) show, CARs are useful in event studies if the date of event is not very precise, because returns are accumulated over time intervals that contain actual event days. In fact, a major use of cumulative abnormal return (CAR) is to find out how announcements affect firms with time-series properties of daily data. That is, when the aim is to test whether the sample company continues to receive abnormal returns, the CARs are considered appropriate

(Dissanaike & Le Fur, 2003). Because it is obvious that the effect is possibly spread over a few days surrounding the event day, and, not only the day when the event occurs. In addition, it indicates whether the return is influenced before the event. That is, when it showed a dramatically change, we consider whether there might be information leakages before the event day. We here let CAR as one of the variables, which shows the prior 20 days', if possible, cumulative abnormal return before each event. In accordance with MacKinlay (1997), the cumulative abnormal return of security i is during $t_1$ to $t_2$ where $t_1 \leq t_2$. The CAR from $t_1$ to $t_2$ is the sum of the included abnormal returns,

$$CAR_i(t_1, t_2) = \sum_{t=t_1}^{t_2} AR_{i,t} \qquad (4.2.7)$$

$$AR_{it} \sim N(0, \sigma^2(AR_{it}))$$

Since we obtain all the data by each characteristic, we can then focus on the cluster analysis. The clustering algorithm used in this paper is based on a hierarchical method in which pairs of entities are sequentially merged into clusters. Aim to prevent each variables from being affected by the size of the value, we calculate each parameter with $X/\bar{X}$. When we input all the characteristics as the variables into the SPSS, an appropriate method should be chosen. Meanwhile, because the clustering procedure is a criterion of cluster homogeneity, and further evaluates the distance between the variables, squared Euclidean distance will be as the interval measurement. In our empirical research, we concentrate on the following three parts of samples:

1. Overall selected 37 entities for each event from Thomson Reuters DataStream.

2. Clustered samples with 4 variables, which conclude CAPM-Beta, volatility, daily spread, and volume-weighted return.

3. Clustered samples with all variables, which conclude CARs, CAPM-Beta, volatility, daily spread, and volume-weighted return. Details of the outcome for the corresponding dendrograms after clustering are presented in Appendix, which are the 'closest' companies to each other. As a result, we have 5 different groups for each cluster with different variables, which have been reported in the tables below.

### 4.3. Excess Return Estimation due to Announcement of Data Scandal

Afterwards, we measure the impact from this scandal, that is, the difference between actual return and expected return. Expected returns are estimated under the economic model, in which the Market Model will be used in this paper. Brown & Warner (1985) realize that the market model is both well-specified and relatively powerful under very large conditions. The market model was given in equation (3.1.1), which suggests that this particular statistical model has become a common choice for estimating expected returns. The $α_i$ and $β_i$ are the ordinary least squares (OLS) parameters, which are estimates from the regression of $R_{it}$ on $R_{mt}$ over an estimation period ($T_0$) preceding the event, for example, 250 to 50 days prior to the event. Given the Market Model parameter estimates, the abnormal return for each firm i can be measured according to:

$$AR_{it} = R_{it} - E(α_i + β_i R_{mt}) \qquad (4.3.1)$$

where E is the expected return of stock i on day t, which denoted as $α_i+β_i R_{mt}$ here, $R_{mt}$ is the return of the market portfolio on day t and $R_{it}$ is the actual return for equity i on day t.

### 4.4. Model Estimation and Hypothesis Testing

While this empirical study is concerned to infer a particular event comprehensively, we then cumulate the abnormal return in samples to construct a new statistic that reflects the cumulative effect of events. Since we can cumulate in both time and firm dimensions, and there might be correlations between companies, we thus discuss in different cases.

First is the cumulative abnormal return (CAR), which indicates that the firm-related announcement extending to multiple days for each single stock, cause the event might be affected by a time period, which depict overall inferences in a period (MacKinlay, 1997). We then denoted the cumulative abnormal return as follow:

$$CAR(t_1, t_2) = \sum_{t=t_1}^{t_2} AR_{i,t} \qquad (4.4.1)$$

Second, for measuring the overall effect of new information signal through the industry, the researchers always capture the cumulative effect across the related companies in a time period (Kothari & Warner, 1997). That is, we aggregated the securities in cross-section and across time, which means the cumulative average abnormal return will be used. The CAARs from day $t_1$ to day $t_2$ in the event period is given by

$$CAAR(t_1, t_2) = \sum_{t=t_1}^{t_2} CAR_{i,t} = \sum_{i=1}^{N} \overline{AR_{i,t}} \qquad (4.4.2)$$

For each single day t, the cross-sectional arithmetic daily average abnormal return for N securities is calculated by

$$\overline{AR_t} = \frac{1}{N}\sum_{i=1}^{N} AR_{i,t} \quad (4.4.3)$$

Given that the aim of this paper is to analyze if the data leak scandal by Facebook in March, also its scandal-related events in the next 4 months, have the impact to the industry-related companies. As we mentioned previously, we divided all the tech-industry related companies with different variables for each event. Furthermore, we also do the clustering after the final discussed event in July, in order to find out what it would be after Facebook reported his second quarter report, where the market's reaction from the data leakage scandal will be highly reflected in this report. To examine whether the positive and the negative abnormal performance before and after the announcements are statistically different from zero, both two tailed t-test and the Wilcoxon signed-rank test are computed for the calculated Cumulative Average Abnormal Returns (CAARs).

**Parametric and non-parametric test**

The t-test assumes a normal distribution of abnormal returns or cumulative abnormal returns (Warner & Brown, 1980). If there is a dependence on the sample size, and when the sample size increases, the relevant sample distribution resists the standardized normal distribution, then the test statistics are asymptotically normal. As a result, we can partially approximate the t-distribution ($N_{(0,1)}$) by a normal distribution. Patell (1976)report the calculation of each standard deviation taking into account, the different standard deviations of each security is carried out at the period in the estimation window. However, other researchers Marks and Musumeci (2017) consider that Patell test ignore additional return variance. According to Brown and Warner (1985), who employ a simple t-test by using cross-sectional dependence, assuming that abnormal returns are independent across stocks, we apply the standardized cross-sectional variance across sample firms as:

$$t_{CAAR} = \sqrt{N}\frac{CAAR}{S_{CAAR}} \sim N(0,1) \quad (4.4.4)$$

$$S_{CAAR}^2 = \frac{1}{N-1}\sum_{l=1}^{N}(CAR - CAAR)^2 \quad (4.4.5)$$

The test statistics used to estimate the impact on Facebook itself after the fourth event over the event window is:

$$t_{AR_{i,t}} = \frac{AR_{i,t}}{S_{AR}} \sim N(0,1) \quad (4.4.6)$$

$$S^2_{AR_I} = \frac{1}{M_I-2}\sum_{t=T0}^{T1}(AR_{i,t})^2 \qquad (4.4.7)$$

where $M_i$ is the non-missing returns for security i. To estimate the standard deviation, we use the entire estimation period.

In fact, however, we always fail to do the assumption about overall distribution (Warner & Brown, 1980). In case of non-normality of the abnormal returns, the former parametric t-test may be poorly specified. Therefore, we combine the non-parametric test to increase the importance of significant testing. Several studies have examined the power of non-parametric tests like rank test and signed-rank test (Corrado, 1989; Corrado & Zivney, 1992; Kwok & Brooks, 1990). Cowan (1992) investigate the performance of the generalized sign test, which is also considered specified under the null hypothesis. Further, Kolari and Pynnonen (2011) demonstrate a generalized rank test based on the previous studies. Both the sign and the magnitude of the abnormal performance are considered when calculating test statistic in Wilcoxon-signed rank test (Wilcoxon, 1945), which has no assumption on the distribution.

$$Z = \frac{W - N(N-1)/4}{\sqrt{N(N+1)(2N+1)/24}} \qquad (4.4.8)$$

## 5. Result

In this section, we will discuss the result from the previous study. Meanwhile, the results will be explained event by event, where the cumulative average abnormal returns for each of the four events are illustrated in Figs. 1-4 by plotting the CAAR's from market model. Besides, these samples are further subdivided into sub-samples, in accordance with the different variables associated with several days before each announcement. Within each sample category and sub-sample category, separate plots representing different clusters, where the red plot in the graphic represents the full sample of common stocks, were collected from Thomson Reuters DataStream; the blue plot represents the samples we clustered with 4 mentioned variables, in accordance with their spread and relative risk, and the green plot represents the samples we clustered with 5 variables, respectively. Furthermore, the whole time period during each event will be described, after which this whole period will be divided into two periods: a pre-event period from -5 till -1 and the post-event period from 0 till +20. When there are fewer than 20 days in post-event period, we take all days in this period account. The discussion will be presented in three stages. First, the calculated daily CAARs results will be presented by comparing the cross-sectional and time–series behavior from each cluster in an around event-announcement period, which is (-5, +20) in Figs, 1-4. Next we illustrate whether the data leakage scandal and the later relevant events had impacts across our chosen U.S. tech stocks, in which the pre-and post-announcement period will be discussed separately. Analyzing this impact is essentially a test of our calculated daily cumulative average abnormal returns. Finally, we make a brief comparison of why there might be different results between different clusters.

### 5.1. The Effect of Data Leakage Scandal on Certain Clusters across U.S. Tech Stock Market

We first analyze the impact of the exposure of the scandal from Facebook, which announced that a third-party was allowed to collect user data. It could request access to the user's name, gender, location, time zone and more. They then shared these numerous different kind of data with Cambridge Analytica, which later utilized the data for political campaigns, especially in the U.S. 2016 presidential election and the Brexit vote. After that, Facebook lost 50 billion dollars in market value after the scandal without doubt when the market open on March 19, 2018. In order to find out whether this sudden scandal would also affect the U.S. tech stock markets, we have already presented three certain clusters for this event.

The results for each cluster in this event are given by plots in Figure 1. The number of the samples in each cluster are 37, 28 and 3 respectively. The results here are concerned with the time period of -5 till +14 due to the smooth curve where all three plots are relative stable at zero or, rather, greater than zero from day -5 to day -1.This means all of the sample firms are kept at a steady level and, there were also no other influential events during this time period in the U.S. tech industry. When the data leakage scandal from Facebook was first announced, the CAAR graph (below) shows a fluctuation at day 0. As we can see in Figure 1, the CAARs across the all 37 sample firms with red plot has a positive upward trend of 0.0068 on day 0 and continuous increase till day 3 of 0.0112, after which it immediately goes down till day 14. For the blue plot, which contains 28 sample firms, presents only a slight increase on day 0. Furthermore, it can be seen that the CAARs are relatively stable at zero during the whole time period from day +1 to day +14. In contrast, the CAARs across 3 sample firms with green plot, which are clustered with 5 variables, show a negative downward trend of -0.0113 on day 0. Three days after the Facebook scandal, which is demonstrated from day +1 to day +3, the result is around 0 again, whereas it drops till day 5 to -0.0256. However, the cumulative abnormal return rises to around -0.0079 at day 9 and a couple days later, it shows a negative downward trend again.

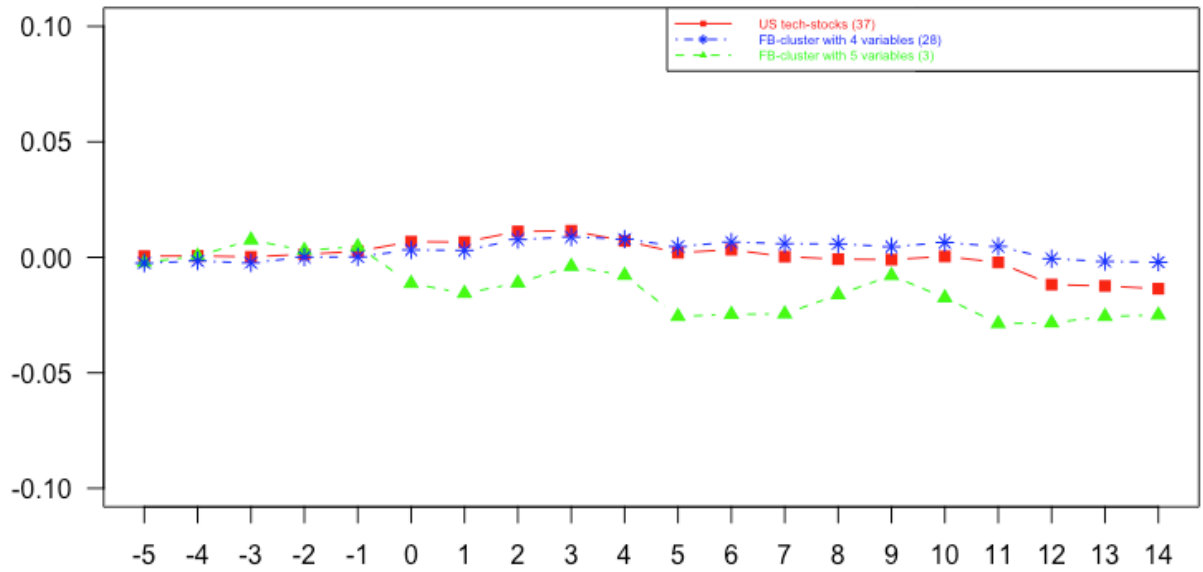

Figure 1. Cumulative Average Abnormal Return for Sample Firms in Event 1

Accordingly, we are concerned with the conclusion of the statistically significant tests of daily CAARs calculated by market model in each time period. As we mentioned earlier, pre- and post-announcement windows are defined for purpose of studying trading behavior before and after news events, respectively. Therefore, we analyze the two time periods (-5, -1) and (0, +14) separately here.

Here it can be seen in Table 2 that all three categories are reported. First of all, in these all 5 days prior to the scandal, it shows that there are no significant CAARs in any of the category except the one from day -5 to -3 in category 3 with parametric test, which is the FB-cluster subsample of firms with 5 variables. Considering that the T-value of 6.3 is not significantly different from zero at a 1% level. Meanwhile, the CAARs in the whole pre-event window (-5, -1) have neither statistical significance on parametric test, nor on the non-parametric test in any time of period. We, therefore, regard this result as consistent with our findings in Figure 1. This conclusion becomes apparent that there is no leakage of the Facebook data leakage scandal before the event date 0, nor other events affect the stock price of firms for the U.S. technology companies. Secondly, we analyze the results after the announcement, which occurred in the post-announcement period. For all 37 sample firms on day 0, the CAAR is greater than zero, where it also indicates a significant value by both parametric test and Wilcoxon test at 10% level at least. Furthermore, by cumulating the average abnormal return from day 0, it reaches 0.0089 at day 3 and shows an obvious positive statistics at 5% level at least by the two tests. However, when a long-term effect is taken into consideration, the results show a non-significant value in non-parametric test among this long time periods. Similar results are observed in samples of category 2, a subsample which consists of 28 firms clustered by 4 variables from this event. It is probably because the investors are unable to define whether it happened among the whole industry or for an individual company at the beginning and, because of the diversity of operating directions in technology industry, it even has a positive effect across the entire tech industry. Several days later, when Facebook was put in the spotlight, the negative effect on other firms might have decreased. Therefore, the overall stock returns of other listed companies in the technology industry performed poorly and were infected by negative news from Facebook. The results, therefore, demonstrate that the H0-1 should be rejected because there are significant test values of CAARs in the event period from day 0 to day +3. As regards the H0-2, when the sub-sample in category 2 is clustered with three risk-focused variables, it cannot be established as well. Examining category 3, however, reported CAARs from day 0 is not much different from zero than any other categories and, rather, smaller than zero. In addition, it decreased for a long time period. Nevertheless, the results have failed to find out the significance in any of the significant test. Also, the H0-2 in this subsample, which is clustered generally across the all variables, cannot be rejected at any significant level. One possible reason for this result could be the small range of portfolios in this sub-sample, it will result in a small T-value in such a situation. Furthermore, as we can see in Appendix (Table B1), the firms in a Facebook-cluster in the first cluster level are Intuit Inc., Workday, together with Facebook. As is well known, Intuit is one of a supplier that provides the financial, accounting, and tax preparation software and related services. Workday, Inc. provides cloud ERP computing software, which concentrates on system for finance, human capital management, and planning. That is, their main businesses seem different rather than being consistent with each other.

Table 2: Daily CAAR of Tech industry sample firms in event 1 and their significant values

|  | All Samples | | | Subsample (with 4 variables) | | | Subsample (with 5 variables) | | |
|---|---|---|---|---|---|---|---|---|---|
| day | CAAR1 | T-test1 | Wilcoxon-test1 | CAAR2 | T-test2 | Wilcoxon-test2 | CAAR3 | T-test3 | Wilcoxon-test3 |
| -5 | 0.0005 | 0.1735 | -0.9127 | -0.0025 | -1.2324 | -0.9127 | -0.0022 | -0.8607 | -1.0690 |
| -4 | 0.0006 | 0.1927 | -0.8524 | -0.0016 | -0.5802 | -0.8524 | 0.0005 | 0.1217 | 0.0000 |
| -3 | 0.0002 | 0.0502 | -0.9278 | -0.0024 | -0.7954 | -0.9278 | 0.0075 | 6.3016** | 1.6036 |
| -2 | 0.0013 | 0.3470 | -0.7166 | 0.0001 | 0.0280 | -0.7166 | 0.0030 | 0.3718 | 0.0000 |
| -1 | 0.0026 | 0.6310 | -0.4149 | 0.0002 | 0.0459 | -0.4149 | 0.0046 | 0.3274 | 0.0000 |
| day | CAAR1 | T-test1 | Wilcoxon-test1 | CAAR2 | T-test2 | Wilcoxon-test2 | CAAR3 | T-test3 | Wilcoxon-test3 |
| 0 | 0.0042 | 1.9876* | 2.3761** | 0.0030 | 1.1853 | 1.7079* | -0.0159 | -1.1377 | -1.0690 |
| 1 | 0.0040 | 1.3301 | 1.9235* | 0.0028 | 0.7686 | 1.5485 | -0.0202 | -0.7911 | -0.5345 |
| 2 | 0.0086 | 2.5624** | 2.9645*** | 0.0075 | 1.9581* | 2.6870*** | -0.0157 | -0.7768 | -0.5345 |
| 3 | 0.0089 | 2.4194** | 2.4063** | 0.0086 | 1.9967* | 2.4138** | -0.0085 | -0.3830 | 0.0000 |
| 4 | 0.0045 | 1.0024 | 1.1692 | 0.0077 | 1.6684 | 1.7306* | -0.0124 | -0.5767 | -0.5345 |
| 5 | -0.0006 | -0.0963 | 0.4903 | 0.0045 | 0.7773 | 1.2297 | -0.0302 | -0.9995 | -1.0690 |
| 6 | 0.0007 | 0.1121 | 0.5205 | 0.0064 | 0.9456 | 1.2069 | -0.0293 | -0.7605 | -0.5345 |
| 7 | -0.0023 | -0.3662 | 0.0075 | 0.0057 | 0.8745 | 1.1158 | -0.0291 | -0.9141 | -1.0690 |
| 8 | -0.0034 | -0.5455 | -0.1584 | 0.0056 | 0.9155 | 1.0019 | -0.0208 | -0.8708 | -1.0690 |
| 9 | -0.0036 | -0.5778 | -0.1131 | 0.0044 | 0.7052 | 0.7742 | -0.0125 | -0.5304 | -0.5345 |
| 10 | -0.0022 | -0.3329 | 0.0679 | 0.0062 | 0.9256 | 1.0930 | -0.0221 | -0.9286 | -1.0690 |
| 11 | -0.0048 | -0.7064 | -0.2640 | 0.0045 | 0.6222 | 0.8881 | -0.0333 | -1.1192 | -1.0690 |
| 12 | -0.0144 | -1.8650* | -1.5463 | -0.0008 | -0.1178 | -0.0683 | -0.0330 | -1.5063 | -1.0690 |
| 13 | -0.0150 | -1.8059* | -1.4257 | -0.0020 | -0.2696 | -0.1594 | -0.0302 | -1.4296 | -1.0690 |
| 14 | -0.0162 | -1.7738* | -1.3351 | -0.0023 | -0.2808 | -0.0455 | -0.0296 | -1.2364 | -1.0690 |

\*\*\* denotes significant at 0.01 level
\*\* denotes significant at 0.05 level
\* denotes significant at 0.1 level

### 5.2. The Effect of Congressional Hearing on Certain Clusters across U.S. Tech Stock Market

After the exposure of the devastating event, Mark Zuckerberg had apologized for their data misuse in public. On April 10, 2018, he was asked to testify by U.S. Senators before Congress over Facebook's user privacy policies and data leakage. It was all about how the social media giant will protects its users. During the testimony, the stock price by Facebook even achieved a growth to 4.5%, which was its biggest one-day gain in nearly two years. We, therefore, discuss here whether the grilling had an effect on the U.S. tech stock market.

Figure 2 depicts the tendency of CAARs across sample firms in three different clusters. It can be seen that the red plot in this graph, which is all 37 sample firms, had a decreasing tendency from day -5 to day -1 of 0.0015 to -0.0125 in the pre-event period. From day 0, when Zuckerberg testified in the U.S. Congress, the CAAR started to show a positive escalating trend till day +2. This means that this event would have an overall effect on the tech stocks market in the U.S. on the event date and also the following short period of time. It tumbled to its lowest level till day +7 of -0.034. Such a result is similar with the blue plot, which is a subsample, and contains 11 firms in total. However, it displays a more negative dramatic downward trend during the post-event period, which reaches -0.0795 on day +7. Moreover, as regards the green plot in this graph, the plot is smooth in the pre-event window and showed an increase trend with fluctuation till day +2 in the post-event window. Since then, the CAARs turned into steady around the zero. From day +7, it started to decline immediately and went below zero.

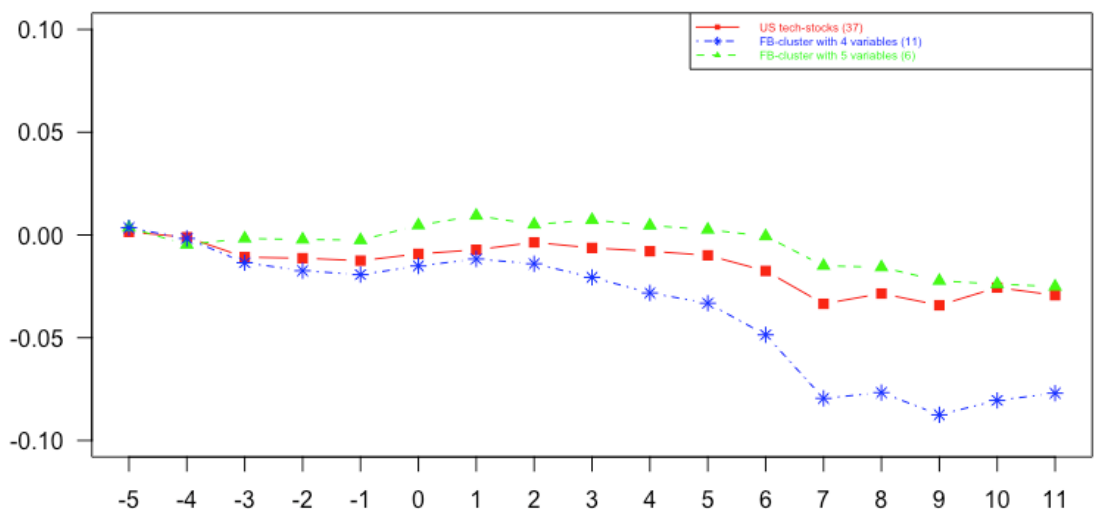

Figure 2. Cumulative Average Abnormal Return for Sample Firms in Event 2

Looking at Table 3, the results of the statistical significant test of CAARs in both pre-event period (-5, -1) and in post-event period (0, +11) are displayed. For all the 37 sample firms, it displays a negative significant effect from t=-5 to t=-1, with a significant value of -2.66 and -2.45 at 5% level in parametric test and Wilcoxon test respectively. On congressional hearing day 0, results for the significant test, show an overall effect across the total U.S. tech industry under the parametric test, whereas the values during the period of (0, +2) in term of both tests were all significantly positive, for 1% level (***). Moreover, the magnitudes of CAARs of the until day +9 are negative and significantly different from zero at 5% level with -2.66 for parametric test and -2.23 with non-parametric test respectively.

Table 3: Daily CAAR of Tech industry sample firms in event 2 and their significant values

| | All Samples | | | Subsample (with 4 variables) | | | Subsample (with 5 variables) | | |
|---|---|---|---|---|---|---|---|---|---|
| day | CAAR1 | T-test1 | Wilcoxon-test1 | CAAR2 | T-test2 | Wilcoxon-test2 | CAAR3 | T-test3 | Wilcoxon-test3 |
| -5 | 0.0015 | 1.0658 | 0.9580 | 0.0035 | 2.0151* | 1.6893* | 0.0036 | 1.6659 | 1.3628 |
| -4 | -0.0011 | -0.5435 | -0.6261 | -0.0017 | -0.5870 | 0.0000 | -0.0047 | -0.9859 | -1.1531 |
| -3 | -0.0108 | -3.1226*** | -2.7985*** | -0.0135 | -5.2103*** | -2.9341*** | -0.0017 | -0.6349 | -0.3145 |
| -2 | -0.0113 | -2.8410*** | -2.8136*** | -0.0174 | -3.5292*** | -2.5784*** | -0.0022 | -0.4290 | -0.5241 |
| -1 | -0.0125 | -2.6572** | -2.4515** | -0.0193 | -3.3811*** | -2.4895** | -0.0025 | -0.4340 | -0.3145 |
| day | CAAR1 | T-test1 | Wilcoxon-test1 | CAAR2 | T-test2 | Wilcoxon-test2 | CAAR3 | T-test3 | Wilcoxon-test3 |
| 0 | 0.0033 | 2.0030** | 1.5463 | 0.0042 | 1.4813 | 1.1558 | 0.0072 | 1.8102 | 1.3628 |
| 1 | 0.0053 | 2.9933*** | 2.8438*** | 0.0076 | 2.2380** | 2.2228** | 0.0119 | 2.1325* | 1.9917** |
| 2 | 0.0089 | 3.5485*** | 2.9192*** | 0.0053 | 2.1288* | 1.7782* | 0.0076 | 2.3147* | 1.7821* |
| 3 | 0.0062 | 2.0061** | 1.8632* | -0.0013 | -0.2563 | -0.2667 | 0.0098 | 1.5499 | 1.3628 |
| 4 | 0.0046 | 1.3401 | 1.2597 | -0.0089 | -1.4784 | -1.3337 | 0.0071 | 1.1233 | 0.9435 |
| 5 | 0.0026 | 0.7246 | 0.7166 | -0.0139 | -2.6752** | -2.2228** | 0.0051 | 0.6895 | 0.3145 |
| 6 | -0.0050 | -1.1563 | -0.7166 | -0.0293 | -4.8334*** | -2.7562*** | 0.0019 | 0.2601 | 0.5241 |
| 7 | -0.0210 | -3.0449*** | -2.4063** | -0.0602 | -5.3843*** | -2.8451*** | -0.0125 | -0.7663 | -0.7338 |
| 8 | -0.0160 | -2.2937** | -1.8632* | -0.0574 | -4.8057*** | -2.7562*** | -0.0132 | -0.6647 | -0.7338 |
| 9 | -0.0217 | -2.6567** | -2.2252** | -0.0682 | -5.0299*** | -2.8451*** | -0.0198 | -0.8729 | -0.7338 |
| 10 | -0.0131 | -1.4225 | -1.5313 | -0.0612 | -4.6540*** | -2.7562*** | -0.0214 | -0.9140 | -0.7338 |
| 11 | -0.0168 | -1.8161* | -1.9386* | -0.0577 | -3.7681*** | -2.6673*** | -0.0226 | -0.9375 | -0.7338 |

\*\*\* denotes significant at 0.01 level
\*\* denotes significant at 0.05 level
\* denotes significant at 0.1 level

Hence, we can consider that the H0-1 should be rejected because of the significant values during day 0 to day +2. In addition, CAARs for the 4-variable clustered subsample contains 11 firms among all the samples, and represents a similar significant value with the whole sample not only in pre-event period from day -5 to day -1, which are -3.38 at 1% level with parametric test and -2.49 at 5% level with non-parametric test respectively, but also at the first two days after the event. However, the overall effect took a few days longer in this subsample where it has a negative downward trend of -0.0577 on day +11 and has a significantly negative parametric value of -3.77 and non-parametric value of -2.67 both at 1% level till day +11. This is obvious consistent with the fluctuation of 17 days' cumulative abnormal average returns in Figure 2. The conclusion convinces us that the H0-2, when the sub-sample here was clustered with 4 variables, is invalid as well. One explanation for this finding might be the uncertainty of the grilling prior to the event day. When it was reported by the public that Facebook will be argued to testify before Congress over the data misuse scandal by answering tough questions, the market were in panic and did not know what will happen on the day of the testimony. Therefore, it shows a negative effect in the pre-event window. The market's reaction become unexpected when the situation became clear and brighter on the event day. Along with the upcoming release of the first quarter earnings results in the following days, the market messed it up again. We conclude that this event had a contagion effect on corresponding subsample firms. The results in another subsample, which are clustered with 5 variables, contradicts the results to the others as they reacts no significant effect at a 5% level both before and after the event in general. Therefore, the effect on 5-certain-variables cluster in the U.S. tech stocks markets may not have the statistical power to reject the null hypothesis.

## 5.3. The Effect of the First Quarterly Report on Certain Clusters across U.S. Tech Stock Market

It should be noted that the stock price for Facebook on April 26, 2018, which opened more than 8 percent higher a day after reporting their first quarter earnings. A large body of empirical studies on financial reports has emphasized that investors response to the release of accounting earnings information by driving stock prices up or down in accordance with the sign of unexpected earnings (Ball & Brown, 1968). Apparently, the variability of stock returns is greater in the time of firms' interim earnings announcement than in non-announcement periods. This is because the market receives more information than other times during the earnings report release, on an average (May, 1971).

The upper panel of Figure 3 plots CAARs for the 37 sample firms in time period (-5, +20). It shows a fluctuation in the time period (-5, -1), which is similar to the other two subsamples, whereas they show a further downward trend compared to the red plot. For example, the pre-event CAARs for the blue plot from t = -5 to t = -1 accumulate from -0.016 to -0.026. Despite the stock price of Facebook had a substantial rise on the event day 0, the CAARs in three categories do not demonstrate a great swing at that day. After the event day, the red plot had a stable fluctuation below the zero even though it had a sudden drop on day +11. In addition, although the CAARs in green plot are lower than the red one, it shows a similar trend with the red one. Evidence of 4-variables-clustered subsample with blue plot in the graph displays a sharp decline from day 0 of -0.022 to day +4 of -0.057 in a post-event period. After day +4, CAARs start to grow slowly.

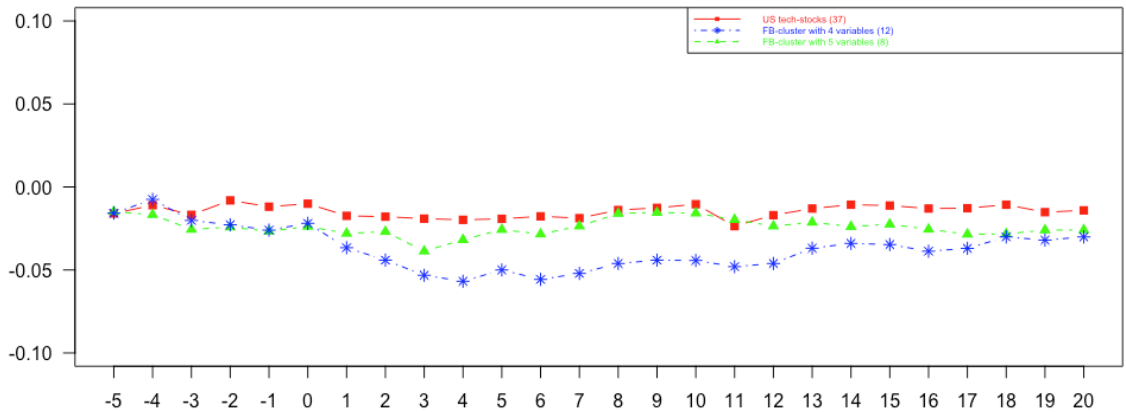

Figure 3: Cumulative Average Abnormal Return for Sample Firms in Event 3

This pattern is consistent with the results in Table 4. From day -5 to day -1 before the quarterly financial report, both subsamples with 4-variables and 5-variables show a negative trend with significant value at 1% and 5% level with parametric test respectively. In consideration of the CAARs for the whole sample firms, it shows negative significant values during the time period (-5, -3) at the 1% level with parametric test, while the portfolio test statistics during (-5, -1) have only a slight significant value of -1.79 and -1.85 at the 10% level for parametric and non-parametric tests respectively. It is probably because the markets become unstable before the announcement of the quarterly report. They do not exactly know the tendency for the earnings. Surprisingly, the results do not show great significant values in the post-event period. As shown in the 37-sample category, cumulative average abnormal returns are almost below zero in a short post-announcement period. No significant results appears with parametric test, while the drop CAARs drops on day +1, +2, +3 and +4 are -0.0055, -0.006, -0.0072, and -0.0079 with significant Wilcoxon T-value of -2.04**, -1.74*, -1.83*, and -1.86*. The result indicates that the financial report from Facebook had only a slight impact across all the sample firms at a 10% level in a short-term. According to Corrado & Zivney (1992), no matter how variance increased or how cross-sectional distribution of event date skewed, rank and sign tests seem less affected and are more valid, unless for a more than 100-day CAR window. It is further demonstrated by Kolari and Pynnönen (2010), who indicate that Wilcoxon (1945) signed-rank test has relatively higher ability in performance than parametric tests, especially for a fat-tailed distribution. Only if, there is no increase in event-induced volatility, both parametric and nonparametric tests are approximately equally powerful (Kolari & Pynnönen, 2010).

Table 4: Daily CAAR of Tech industry sample firms in event 3 and their significant values

| | All Samples | | | Subsample (with 4 variables) | | | Subsample (with 5 variables) | | |
|---|---|---|---|---|---|---|---|---|---|
| day | CAAR1 | T-test1 | Wilcoxon-test1 | CAAR2 | T-test2 | Wilcoxon-test2 | CAAR3 | T-test3 | Wilcoxon-test3 |
| -5 | -0.0160 | -4.4270*** | -3.6433*** | -0.0158 | -2.8937** | -2.2749** | -0.0149 | -2.3077* | -1.6803* |
| -4 | -0.0110 | -2.7291*** | -2.2554** | -0.0075 | -1.3502 | -1.1767 | -0.0167 | -1.7386 | -1.5403 |
| -3 | -0.0168 | -3.1453*** | -2.5571** | -0.0202 | -2.3499** | -1.8827* | -0.0256 | -2.3157* | -1.8204* |
| -2 | -0.0081 | -1.2658 | -1.3804 | -0.0228 | -2.6888** | -2.3534** | -0.0242 | -2.8510** | -2.1004** |
| -1 | -0.0119 | -1.7897* | -1.8481* | -0.0260 | -3.3179*** | -2.5103** | -0.0267 | -3.4320** | -2.1004** |

| day | CAAR1 | T-test1 | Wilcoxon-test1 | CAAR2 | T-test2 | Wilcoxon-test2 | CAAR3 | T-test3 | Wilcoxon-test3 |
|---|---|---|---|---|---|---|---|---|---|
| 0 | 0.0018 | 0.4612 | -1.3201 | 0.0041 | 0.6908 | -0.2353 | 0.0029 | 0.3149 | -0.9802 |
| 1 | -0.0055 | -1.0066 | -2.0442** | -0.0106 | -1.0323 | -1.4120 | -0.0013 | -0.1301 | -1.1202 |
| 2 | -0.0060 | -1.0514 | -1.7425* | -0.0181 | -1.4613 | -1.8043* | -0.0002 | -0.0238 | -0.7001 |
| 3 | -0.0072 | -1.0886 | -1.8330* | -0.0272 | -1.8115* | -2.3534** | -0.0120 | -0.6718 | -1.1202 |
| 4 | -0.0079 | -1.0636 | -1.8632* | -0.0308 | -1.7430 | -2.2749** | -0.0051 | -0.2438 | -0.5601 |
| 5 | -0.0073 | -0.9244 | -0.8825 | -0.0240 | -1.3315 | -1.2551 | 0.0010 | 0.0476 | 0.1400 |
| 6 | -0.0058 | -0.6994 | -0.6713 | -0.0297 | -1.5381 | -1.4905 | -0.0017 | -0.0750 | 0.0000 |
| 7 | -0.0069 | -0.7809 | -0.8524 | -0.0261 | -1.2976 | -1.4120 | 0.0031 | 0.1285 | 0.0000 |
| 8 | -0.0020 | -0.2100 | 0.2489 | -0.0201 | -0.9199 | -0.5491 | 0.0107 | 0.4393 | 0.5601 |
| 9 | -0.0007 | -0.0758 | 0.3093 | -0.0181 | -0.8170 | -0.5491 | 0.0113 | 0.4554 | 0.4201 |
| 10 | 0.0015 | 0.1439 | 0.9127 | -0.0183 | -0.7889 | -0.3922 | 0.0109 | 0.4021 | 0.5601 |
| 11 | -0.0117 | -0.7806 | 0.2791 | -0.0220 | -0.9038 | -0.4707 | 0.0070 | 0.2435 | 0.5601 |
| 12 | -0.0051 | -0.3742 | 0.5205 | -0.0201 | -0.8724 | -0.5491 | 0.0031 | 0.1075 | 0.2801 |
| 13 | -0.0011 | -0.0847 | 0.5808 | -0.0109 | -0.5301 | -0.1569 | 0.0054 | 0.2003 | 0.1400 |
| 14 | 0.0012 | 0.0967 | 0.8222 | -0.0080 | -0.4628 | -0.2353 | 0.0028 | 0.1122 | -0.2801 |
| 15 | 0.0007 | 0.0562 | 0.6713 | -0.0087 | -0.4652 | -0.3138 | 0.0042 | 0.1610 | -0.1400 |
| 16 | -0.0011 | -0.0793 | 0.3847 | -0.0127 | -0.5669 | -0.1569 | 0.0012 | 0.0390 | 0.1400 |
| 17 | -0.0009 | -0.0624 | 0.3998 | -0.0110 | -0.5039 | -0.0784 | -0.0018 | -0.0577 | -0.1400 |
| 18 | 0.0012 | 0.0828 | 0.6563 | -0.0041 | -0.1925 | 0.4707 | -0.0016 | -0.0530 | -0.1400 |
| 19 | -0.0033 | -0.2199 | 0.4149 | -0.0060 | -0.2760 | 0.4707 | 0.0007 | 0.0207 | 0.0000 |
| 20 | -0.0022 | -0.1367 | 0.3545 | -0.0041 | -0.1780 | 0.3138 | 0.0008 | 0.0234 | 0.0000 |

\*\*\* denotes significant at 0.01 level  
\*\* denotes significant at 0.05 level  
\* denotes significant at 0.1 level

Hence, because of the short period of our test window, we prefer the outcome by the Wilcoxon signed-rank test. As a result, this finding implies that the H0-1 can be rejected at approximately the 10% significance level of the test. Meanwhile, the subsample with 4 variables depicts a similar result. CAARs during the period (0, +4) of -0.0308 interpret strongly in the results of Wilcoxon's nonparametric test with a negative significant value of -2.27 at 5% level. In fact, compared to the CAARs from day 0 to day +4 for these two categories, the market response weaker to the announcement for a longer period from day 0 to day +20. In general, when the data misuse scandal was occurred on March, it did not affect the financial data for Facebook itself that much. It was a great achievement in its business at the beginning of 2018. It should be noted that this study, however, focuses on the impact across a whole industry. Not only Facebook, but also other technical firms are reporting their financial reports during this time period. Therefore, such results will occur. In other words, this is contrary to hypothesis that a first quarterly report from Facebook had no overall effect on the 4-variables clustered U.S. tech stocks market. Contrary to the results for the other two categories, the parametric t-test as well as the nonparametric Wilcoxon signed-rank test statistics, however, suggest that the first-quarterly report from Facebook had no overall significant impact on the 5-variables cluster in tech market because there are no significant values during the post-event period (+1, +20). Thus, H0-2 can be confirmed for 5-variables subsample, but not for 4-variables subsample. One reason is that the earnings and income in the first quarter is better than expected. The number of daily active users in Facebook indicates that though it was influenced by the misuse of data scandal, the company still continues to develop steadily.

**5.4. The Effect of Second Quarterly Report on Certain Clusters across U.S. Tech Stock Market**

For the fourth event, which considered a release of the quarterly report by Facebook as well, there was a drastic slump in the stock price on July 26, 2018. Obviously, it was a financial report from the second quarter, during which there was the great spoken scandal. As a result, the poor performance in relation to the financial target in this quarter did have an influence on the stock price of Facebook itself on that day. we therefore measured the cumulative average abnormal returns in purpose of finding out the impact on certain clusters in the U.S. tech stock market.

Interestingly, the fluctuation in CAARs for all three plots in the whole (-5, +20) period shown in Figure 4 is almost the same. In the pre-event period of 5 days, the CAARs are continuous, falling around to -0.03 for all the three categories. However, they show an upward tendency on the event day 0. Since then, they still maintain the same trend, slightly decreasing till day +20. This outcome is consistent with the findings of Firth (1976), whose investigation found that company's earnings report announcement will affect the stock price behavior of other firms in the same industry.

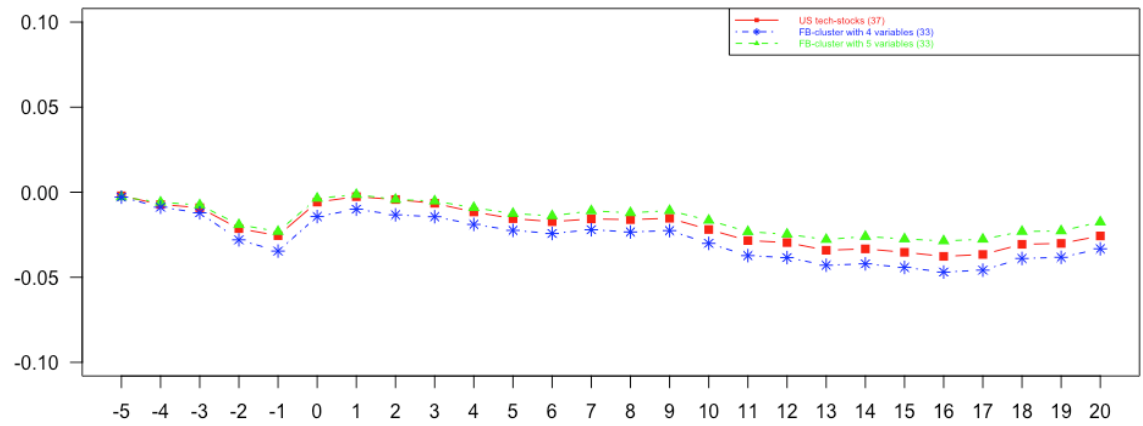

iv

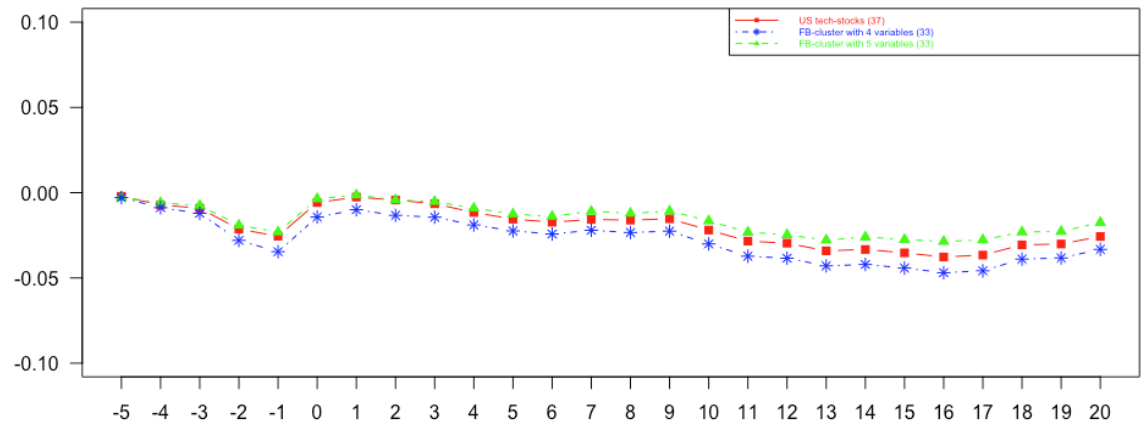

Figure 4. Cumulative Average Abnormal Return for Sample Firms in Event 4

This supports the test results, and the data also disclose that the event is strongly and negatively correlated to the pre-event CAARs. In this event, there are 37, 33 and 33 sample firms in each category. Nevertheless, the 33 firms in both two subsamples are different from each other. As we can see in Table 5, not only in the t-test, but also the Wilcoxon signed-rank test shows a highly negative significant value at 1% level from day -5 to day -1 for all three categories, which correspond to the previous results of CAARs. The reason for this is that in the period before the quarterly report announced, the investors have already expected this report as bad news. Therefore, the overall stock returns of other public companies in the technology industry have performed poorly because of the negative expectations on the part of investors, which indicates that they have been infected by negative news from Facebook. The value of CAARs for the whole 37 sample firms on the announcement day 0 is 0.0197 with Wilcoxon T-value of 4.25*** while the parametric t-test value is at 5% level.

Table 5: Daily CAAR of Tech industry sample firms in event 4 and their significant values

| | All Samples | | | Subsample (with 4 variables) | | | Subsample (with 5 variables) | | |
|---|---|---|---|---|---|---|---|---|---|
| day | CAAR1 | T-test1 | Wilcoxon-test1 | CAAR2 | T-test2 | Wilcoxon-test2 | CAAR3 | T-test3 | Wilcoxon-test3 |
| -5 | -0.0021 | -1.1830 | -0.9429 | -0.0029 | -1.5489 | -1.3133 | -0.0028 | -1.5333 | -1.3848 |
| -4 | -0.0072 | -2.6815** | -2.1649** | -0.0088 | -3.1098*** | -2.5462** | -0.0059 | -2.4987** | -2.0459** |
| -3 | -0.0091 | -2.3915** | -2.2403** | -0.0121 | -3.1761*** | -2.7784*** | -0.0075 | -2.4015** | -2.1173** |
| -2 | -0.0214 | -3.5996*** | -3.1304*** | -0.028 | -5.0348*** | -3.9041*** | -0.0191 | -3.4765*** | -2.9571*** |
| -1 | -0.0255 | -3.3763*** | -2.9192*** | -0.0346 | -5.0289*** | -3.8863*** | -0.0231 | -3.3132*** | -2.7606*** |
| day | CAAR1 | T-test1 | Wilcoxon-test1 | CAAR2 | T-test2 | Wilcoxon-test2 | CAAR3 | T-test3 | Wilcoxon-test3 |
| 0 | 0.0197 | 2.4660** | 4.2468*** | 0.0203 | 2.2651** | 3.9041*** | 0.0195 | 2.1838** | 3.9041*** |
| 1 | 0.0229 | 2.2847** | 2.8438*** | 0.0246 | 2.1996** | 2.6534*** | 0.0217 | 1.9309* | 2.3318** |
| 2 | 0.0212 | 1.8476* | 1.9084* | 0.0213 | 1.6609 | 1.5992 | 0.0188 | 1.4687 | 1.2954 |
| 3 | 0.0190 | 1.6481 | 1.4257 | 0.0202 | 1.5655 | 1.2954 | 0.0179 | 1.3813 | 0.9559 |
| 4 | 0.0139 | 1.2445 | 1.0032 | 0.0157 | 1.2556 | 0.9917 | 0.0140 | 1.1164 | 0.7951 |
| 5 | 0.0100 | 0.8892 | 0.5808 | 0.0122 | 0.9716 | 0.8130 | 0.0106 | 0.8393 | 0.5092 |
| 6 | 0.0082 | 0.7013 | 0.2942 | 0.0103 | 0.7925 | 0.5628 | 0.0092 | 0.7040 | 0.3127 |
| 7 | 0.0098 | 0.8414 | 0.3244 | 0.0125 | 0.9655 | 0.7236 | 0.0121 | 0.9370 | 0.5628 |
| 8 | 0.0094 | 0.7984 | 0.6563 | 0.0112 | 0.8584 | 0.7594 | 0.0110 | 0.8433 | 0.6700 |
| 9 | 0.0102 | 0.8497 | 0.7769 | 0.0120 | 0.8941 | 0.8487 | 0.0123 | 0.9152 | 0.8130 |
| 10 | 0.0036 | 0.2858 | 0.3696 | 0.0045 | 0.3222 | 0.4735 | 0.0066 | 0.4737 | 0.6522 |
| 11 | -0.0029 | -0.2335 | -0.1433 | -0.0027 | -0.1887 | -0.0804 | -0.0001 | -0.0071 | 0.1876 |
| 12 | -0.0041 | -0.3048 | -0.2942 | -0.0038 | -0.2518 | -0.1697 | -0.0016 | -0.1072 | -0.0268 |
| 13 | -0.0086 | -0.6222 | -0.5808 | -0.0083 | -0.5385 | -0.4914 | -0.0047 | -0.3129 | -0.1340 |
| 14 | -0.0078 | -0.5302 | -0.3394 | -0.0075 | -0.4579 | -0.2770 | -0.0029 | -0.1875 | 0.0804 |
| 15 | -0.0098 | -0.6585 | -0.5054 | -0.0096 | -0.5774 | -0.3842 | -0.0044 | -0.2776 | -0.1161 |
| 16 | -0.0122 | -0.7931 | -1.0787 | -0.0123 | -0.7142 | -0.8666 | -0.0055 | -0.3394 | -0.5092 |
| 17 | -0.0111 | -0.7098 | -0.9429 | -0.0112 | -0.6411 | -0.7773 | -0.0046 | -0.2722 | -0.3484 |
| 18 | -0.0051 | -0.3260 | -0.4601 | -0.0043 | -0.2495 | -0.3306 | 0.0000 | -0.0001 | -0.0804 |
| 19 | -0.0046 | -0.2886 | -0.3394 | -0.0037 | -0.2028 | -0.2591 | 0.0003 | 0.0199 | 0.0089 |
| 20 | -0.0002 | -0.0112 | -0.1433 | 0.0013 | 0.0677 | -0.0089 | 0.0056 | 0.2933 | 0.2591 |

*** denotes significant at 0.01 level
** denotes significant at 0.05 level
* denotes significant at 0.1 level

Meanwhile, the test results for both subsamples are also significant at the 1% level under the non-parametric test while the parametric t-test value is at 5% level as well. These negative and statistical significant CAARs are interpreted as a strong rejection for both H0-1 and H0-2 in the post-announcement period from t=0 to t=+1. However, the difference in CAARs is not significant at 10%, 5%, or 1% significant level in a longer time period from day 0 to day +20. This is not surprising since the portfolios in each of these samples contain primarily the same firms.

Specifically, we aim to find out whether the cluster will change over time as regards the release of the second quarter report. So we extend our study by doing a further clustering with the next 20 days. That is, the 20 days after the fourth event day, July 26, 2018. As a consequence, there is only Facebook in a cluster in both the 5-variables cluster and 4-variables cluster, which is shown in Appendix (Table B5 & C5). In other words, this report was reflected in the stock market. We then assume that the release of the report might have had a great influence on Facebook itself. By comparing the daily abnormal returns of Facebook, we represent the calculated ARs from day -5 to day +20 in Figure 5. Apparently, in the pre-announcement time period, the daily abnormal return by Facebook was stable around the zero. What is noticeable for the abnormal return on day 0 is that the abnormal return suddenly plummeted to -0.1918, which reached the bottom for a long time period, as expected. The result indicates that Facebook reacted more negatively on day 0 to the proposals of the release of the financial report when it directly correlated to the achievements for the second quarter. What is more, there was also a greater falling off for Facebook itself compared to the sample tech stocks. It indicates that the market reacted more to this announcement when the report showed a great loss in this quarter, in which the entire scandal is contained. As a result, there could have been a significant increase in trading volumes which also inflected the stock price of Facebook (Morse, 1981). From day +1 to day +6, the daily abnormal return shows a slight fluctuation in a few days after the event. Nevertheless, a positive upward trend occurs on day +7 of 0.04, which stands on the peak in the whole (-5, +20) period. After day +8, the daily abnormal return is stable at zero again, though there is an obvious falling off on day +15.

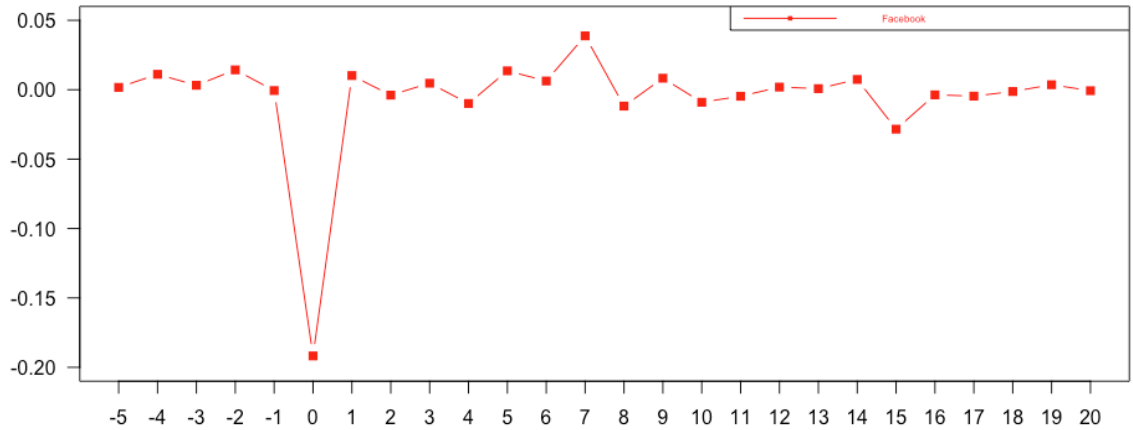

Figure 5. Abnormal Return for FB after Event 4

In Table 6, the daily abnormal return for Facebook is also given in this time period, together with its parametric and non-parametric T-values during this whole period. Here, it shows that from t= -5 to t= -3, there is no significant T-value that is different from zero at a 10% level in the pre-announcement window. Since day -2, though the result in t-test is still not significantly different from zero, the non-parametric shows a slight positive significant value at 10% level. As expected, the parametric T-test is consistent with the plot we had shown before. The only value that is significant is the value at day 0, which is the event day. This is significant where the P-value far below than 1%. This negative and statistically significant AR is -0.1918 on this event day with T-value of -12.7 strongly leads to the interpretation that the release of the report is more correlated with Facebook on heels of the Cambridge Analytica privacy scandal earlier this year. However, in a longer period which is to day +20 in our study, it exhibits the daily AR only in day +7 and in day +15, which are 0.0388 and -0.0284 with T-values of 2.647** and -1.88* respectively. It indicates that the effect is immediately outstanding at the beginning, but in the long run, the effect becomes smaller.

Table 6: Daily AR of Facebook after event 4 and significant values

| day | AR | T-test | Wilcoxon-test |
|---|---|---|---|
| -5 | 0.0017 | 0.1113 | 1.0000 |
| -4 | 0.0111 | 0.7359 | 1.3416 |
| -3 | 0.0032 | 0.2130 | 1.6036 |
| -2 | 0.0143 | 0.9466 | 1.8257* |
| -1 | -0.0006 | -0.0379 | 1.7529* |
| 0 | -0.1918 | -12.6974*** | 0.7338 |
| 1 | 0.0102 | 0.6769 | 1.0142 |
| 2 | -0.0039 | -0.2600 | 0.7001 |
| 3 | 0.0047 | 0.3080 | 1.0070 |
| 4 | -0.0099 | -0.6540 | 0.6625 |
| 5 | 0.0136 | 0.9033 | 0.9780 |
| 6 | 0.0063 | 0.4174 | 1.1767 |
| 7 | 0.0388 | 2.5680** | 1.4327 |
| 8 | -0.0118 | -0.7842 | 1.0358 |
| 9 | 0.0083 | 0.5502 | 1.1927 |
| 10 | -0.0090 | -0.5968 | 0.9308 |
| 11 | -0.0047 | -0.3090 | 0.7811 |
| 12 | 0.0019 | 0.1236 | 0.8492 |
| 13 | 0.0008 | 0.0504 | 0.9256 |
| 14 | 0.0074 | 0.4893 | 1.0826 |
| 15 | -0.0284 | -1.8774* | 0.6778 |
| 16 | -0.0037 | -0.2467 | 0.5681 |
| 17 | -0.0046 | -0.3061 | 0.4562 |
| 18 | -0.0013 | -0.0888 | 0.4286 |
| 19 | 0.0036 | 0.2402 | 0.4978 |
| 20 | -0.0007 | -0.0452 | 0.4953 |

\*\*\* denotes significant at 0.01 level
\*\* denotes significant at 0.05 level
\* denotes significant at 0.1 level

## 6. Conclusion

This paper investigates the impact of the Facebook data leakage scandal and subsequent related events on the value of stocks within the American technology industry, focusing on companies listed on the NASDAQ 100. Through a clustering procedure, we create sub-samples with different firm-specific variables and analyze their CAARs across various time periods following the scandal events.

From an industry-wide perspective, we find significant positive CAARs across sample firms in the U.S. for the three days following the announcement of the scandal and related events. This suggests that the tech industry as a whole did not suffer a negative impact from these events; in fact, other tech-industry firms experienced profits. However, this trend is not observed for event 3, when Facebook reported its first quarterly earnings, which resulted in a significantly negative effect despite the satisfactory report.

When considering subsamples based on specific risk measures, the consequences vary. All four scandal-related events show significant overall effects on their corresponding subsamples from the event day to day +3. However, the impact on certain variable subsamples differs. While the first three events generally have no significant overall impact on the 5-variable cluster in the tech market, the last scandal-related announcements do have a significant effect from day 0 to day +1. Additionally, after Facebook reported second-quarter earnings, only Facebook was clustered in subsamples, indicating a notably negative effect on the company's stock price following the poor earnings announcement.

As discussed in Utz (2019), corporate scandals can have two distinct impacts: exposing internal weaknesses within a company and motivating managers to improve policies. When analyzing the economic effects of one of the largest U.S. privacy scandals, such as the Facebook data leakage scandal, it becomes evident that such firm-specific scandals also significantly affect entire industries. Our findings in this paper underscore the influence of Facebook's scandal on its industry counterparts.

In general, Facebook's dominance in the market suggests it operates as a monopoly. While Google competes for digital ad dollars and Twitter focuses on social services, neither offers the same breadth and depth of services as Facebook. The resources controlled by Facebook vary significantly across different companies. Barney (1991) argues that companies possessing rare, valuable, imperfectly imitable, and non-substitutable resources maintain a competitive advantage. Both tangible and intangible resources contribute to Facebook's position as a market leader and its ability to maintain a competitive edge.

However, Facebook's vast repository of user data, as the world's largest social platform, poses significant risks if not securely managed. The potential fallout from a breach of trust regarding data security could far outweigh the direct consequences of information leakage. In today's internet landscape, emphasis on openness is paramount, with most major internet companies launching open platforms. Ensuring robust oversight of third-party developers to prevent incidents similar to those experienced by Facebook has become an urgent priority for all enterprises.

Considering the limitations regarding the explicit size effect (Schwert 1983) and liquidity (Gomber, Schweickert, & Theissen, 2015), future research is advised to prioritize the identification of market anomalies (Strong, 1992) when selecting samples. Additionally, it is recommended to incorporate a broader range of variables, including various financial indicators, to enhance the comprehensiveness of the analysis.

Moreover, potential confounding events may occur during the period when listed firms release their quarterly reports (McWilliams & Siegel, 1997), thereby influencing the chosen event day. Therefore, it is imperative to exclude firms from the sample on days when such confounding events occur.

Our study introduces an approach that has not been previously explored, where pre-clustering based on firms' specific features can yield more precise analyses and consequently more relevant empirical results. The clustering principle should assist in identifying directly associated companies and thereby mitigating the impact of randomness. This was indeed demonstrated in the case of the key event, "The Effect of Congressional Hearing on Certain Clusters across U.S. Tech Stock Market," which was found to be delayed and significantly negative. Hence, we suggest employing the clustering method when conducting such or similar event studies.

**Appendix**

Table A: All sample firms under the categories of 'Software and Computer Services', and 'Technology Hardware and Equipment'

| U.S. Tech Firms | Code | U.S. Tech Firms | Code |
| --- | --- | --- | --- |
| ADOBE SYSTEMS | ADBE | LAM RESEARCH | LRCX |
| ADVANCED | AMD | MAXIM INTEGRATED PRDS | MXIM |
| ALPHABET A | GOOGL | MICROCHIP TECH. | MCHP |
| ALPHABET C | GOOG | MICRON TECHNOLOGY | MU |
| ANALOG DEVICES | ADI | MICROSOFT | MSFT |
| APPLE | AAPL | NETAPP | NTAP |
| APPLIED MATS. | AMAT | NVIDIA | NVDA |
| AUTODESK | ADSK | NXP SEMICONDUCTORS | NXPI |
| BROADCOM | AVGO | QUALCOMM | QCOM |
| CADENCE DESIGN SYS. | CDNS | SEAGATE TECH. | STX |
| CERNER | CERN | SKYWORKS SOLUTIONS | SWKS |
| CHECK POINT SFTW.TECHS. | CHKP | SYMANTEC | SYMC |
| CISCO SYSTEMS | CSCO | SYNOPSYS | SNPS |
| CITRIX SYS. | CTXS | TEXAS INSTRUMENTS | TXN |
| COGNIZANT TECH.SLTN.'A' | CTSH | VERISIGN | VRSN |
| FACEBOOK CLASS A | FB | WESTERN DIGITAL | WDC |
| INTEL | INTC | WORKDAY CLASS A | WDAY |
| INTUIT | INTU | XILINX | XLNX |
| KLA TENCOR | KLAC | | |

Table B1: All sample firms are clustered in Event 1 with 5 variables

| U.S. Tech Firms | Code |
|---|---|
| FACEBOOK CLASS A | FB |
| INTEL | INTC |
| WORKDAY CLASS A | WDAY |

Table B2: All sample firms are clustered in Event 2 with 5 variables

| U.S. Tech Firms | Code |
|---|---|
| COGNIZANT TECH.SLTN.'A' | CTSH |
| FACEBOOK CLASS A | FB |
| NXP SEMICONDUCTORS | NXPI |
| QUALCOMM | QCOM |
| SYNOPSYS | SNPS |
| VERISIGN | VRSN |

Table B3: All sample firms are clustered in Event 3 with 5 variables

| U.S. Tech Firms | Code |
|---|---|
| ADOBE SYSTEMS | ADBE |
| APPLE | AAPL |
| AUTODESK | ADSK |
| BROADCOM | AVGO |
| FACEBOOK CLASS A | FB |
| MAXIM INTEGRATED PRDS | MXIM |
| NVIDIA | NVDA |
| SEAGATE TECH. | STX |

Table B4: All sample firms are clustered in Event 4 with 5 variables

| U.S. Tech Firms | Code | U.S. Tech Firms | Code |
|---|---|---|---|
| ADOBE SYSTEMS | ADBE | KLA TENCOR | KLAC |
| ADVANCED | AMD | LAM RESEARCH | LRCX |
| ALPHABET A | GOOGL | MAXIM INTEGRATED PRDS | MXIM |
| ALPHABET C | GOOG | MICROCHIP TECH. | MCHP |
| ANALOG DEVICES | ADI | MICROSOFT | MSFT |
| APPLE | AAPL | NETAPP | NTAP |
| APPLIED MATS. | AMAT | NXP SEMICONDUCTORS | NXPI |
| AUTODESK | ADSK | QUALCOMM | QCOM |
| CADENCE DESIGN SYS. | CDNS | SEAGATE TECH. | STX |
| CERNER | CERN | SYMANTEC | SYMC |
| CHECK POINT SFTW.TECHS. | CHKP | SYNOPSYS | SNPS |
| CISCO SYSTEMS | CSCO | TEXAS INSTRUMENTS | TXN |
| CITRIX SYS. | CTXS | VERISIGN | VRSN |
| COGNIZANT TECH.SLTN.'A' | CTSH | WESTERN DIGITAL | WDC |
| FACEBOOK CLASS A | FB | WORKDAY CLASS A | WDAY |
| INTEL | INTC | XILINX | XLNX |
| INTUIT | INTU | | |

Table B5: All sample firms are clustered after Event 4 with 5 variables

| U.S. Tech Firms | Code |
|---|---|
| FACEBOOK CLASS A | FB |

Table C1: All sample firms are clustered in Event 1 with 4 variables

| U.S. Tech Firms | Code | U.S. Tech Firms | Code |
| --- | --- | --- | --- |
| ADOBE SYSTEMS | ADBE | INTUIT | INTU |
| ADVANCED | AMD | KLA TENCOR | KLAC |
| ANALOG DEVICES | ADI | MAXIM INTEGRATED PRDS | MXIM |
| APPLE | AAPL | MICROSOFT | MSFT |
| APPLIED MATS. | AMAT | NETAPP | NTAP |
| BROADCOM | AVGO | NVIDIA | NVDA |
| CADENCE DESIGN SYS. | CDNS | QUALCOMM | QCOM |
| CERNER | CERN | SKYWORKS SOLUTIONS | SWKS |
| CHECK POINT SFTW.TECHS. | CHKP | SYMANTEC | SYMC |
| CISCO SYSTEMS | CSCO | SYNOPSYS | SNPS |
| CITRIX SYS. | CTXS | TEXAS INSTRUMENTS | TXN |
| COGNIZANT TECH.SLTN.'A' | CTSH | VERISIGN | VRSN |
| FACEBOOK CLASS A | FB | WORKDAY CLASS A | WDAY |
| INTEL | INTC | XILINX | XLNX |

Table C2: All sample firms are clustered in Event 2 with 4 variables

| U.S. Tech Firms | Code | U.S. Tech Firms | Code |
| --- | --- | --- | --- |
| APPLIED MATS. | AMAT | NXP SEMICONDUCTORS | NXPI |
| AUTODESK | ADSK | QUALCOMM | QCOM |
| FACEBOOK CLASS A | FB | SKYWORKS SOLUTIONS | SWKS |
| KLA TENCOR | KLAC | TEXAS INSTRUMENTS | TXN |
| MAXIM INTEGRATED PRDS | MXIM | XILINX | XLNX |
| MICROCHIP TECH. | MCHP | | |

Table C3: All sample firms are clustered in Event 3 with 4 variables

| U.S. Tech Firms | Code | U.S. Tech Firms | Code |
| --- | --- | --- | --- |
| ADOBE SYSTEMS | ADBE | MICRON TECHNOLOGY | MU |
| ANALOG DEVICES | ADI | MICROSOFT | MSFT |
| AUTODESK | ADSK | NVIDIA | NVDA |
| BROADCOM | AVGO | SEAGATE TECH. | STX |
| FACEBOOK CLASS A | FB | VERISIGN | VRSN |
| MICROCHIP TECH. | MCHP | WESTERN DIGITAL | WDC |

Table C4: All sample firms are clustered in Event 4 with 4 variables

| U.S. Tech Firms | Code | U.S. Tech Firms | Code |
| --- | --- | --- | --- |
| ADOBE SYSTEMS | ADBE | MICROCHIP TECH. | MCHP |
| ADVANCED | AMD | MICRON TECHNOLOGY | MU |
| ANALOG DEVICES | ADI | MICROSOFT | MSFT |
| APPLE | AAPL | NETAPP | NTAP |
| APPLIED MATS. | AMAT | NVIDIA | NVDA |
| AUTODESK | ADSK | NXP SEMICONDUCTORS | NXPI |
| CADENCE DESIGN SYS. | CDNS | QUALCOMM | QCOM |
| CERNER | CERN | SEAGATE TECH. | STX |
| CISCO SYSTEMS | CSCO | SKYWORKS SOLUTIONS | SWKS |
| CITRIX SYS. | CTXS | SYMANTEC | SYMC |
| COGNIZANT TECH.SLTN.'A' | CTSH | SYNOPSYS | SNPS |
| FACEBOOK CLASS A | FB | TEXAS INSTRUMENTS | TXN |
| INTEL | INTC | VERISIGN | VRSN |
| INTUIT | INTU | WESTERN DIGITAL | WDC |
| KLA TENCOR | KLAC | WORKDAY CLASS A | WDAY |
| LAM RESEARCH | LRCX | XILINX | XLNX |
| MAXIM INTEGRATED PRDS | MXIM | | |

Table C5: All sample firms are clustered after Event 4 with 4 variables

| U.S. Tech Firms | Code |
|---|---|
| FACEBOOK CLASS A | FB |

Table D: Regression Results for Facebook

|  | *Dependent variable:* |
|---|---|
|  | FACEBOOK |
| NASDAQ100 | 1.022*** |
|  | (0.145) |
| Constant | −0.001 |
|  | (0.002) |
| Observations | 52 |
| $R^2$ | 0.498 |
| Adjusted $R^2$ | 0.488 |
| Residual Std. Error | 0.014 (df = 50) |
| F Statistic | 49.580*** (df = 1; 50) |
| *Note:* | *p<0.1; **p<0.05; ***p<0.01 |